\documentclass[journal]{IEEEtran}
\usepackage{graphicx}
\usepackage{epstopdf}
\usepackage{lipsum,amsmath}
\usepackage{subfigure}
\usepackage{diagbox}
\usepackage{array}
\usepackage{cuted}
\usepackage{bm}
\usepackage{multirow}
\graphicspath{ {./} }
\usepackage{multirow}
\usepackage{algorithm}
\usepackage{algorithmic}
\usepackage{color,xcolor}
\usepackage{ulem}
\usepackage{amssymb}    % 对号
\usepackage{bbding}     % 错号
\usepackage[top=2cm, bottom=2cm, left=2cm, right=2cm]{geometry}
\renewcommand\arraystretch{0}
\usepackage{cite}
\ifCLASSINFOpdf
\else
\fi

\begin{document}

% Linebreaks \\ can be used within to get better formatting as desired.
% Do not put math or special symbols in the title.
\title{AIR: Threats of Adversarial Attacks on Deep Learning-Based Information Recovery}

% author names and IEEE memberships
% note positions of commas and nonbreaking spaces ( ~ ) LaTeX will not break
% a structure at a ~ so this keeps an author's name from being broken across
% two lines.
% use \thanks{} to gain access to the first footnote area
% a separate \thanks must be used for each paragraph as LaTeX2e's \thanks
% was not built to handle multiple paragraphs
%

%\author{%Jinyin~Chen,~\IEEEmembership{Member,~IEEE,}
%        Linhui~Ye,~\IEEEmembership{Fellow,~OSA,}
%        and~Jane~Doe,~\IEEEmembership{Life~Fellow,~IEEE}% <-this % stops a space

\author{Jinyin~Chen, 
Jie~Ge, 
Shilian~Zheng*,
Linhui~Ye,
Haibin~Zheng,

Weiguo~Shen,
Keqiang~Yue
and Xiaoniu Yang
% <-this % stops a space

\thanks{Manuscript received XX XX, 2022; revised XX XX, 2022; accepted XX XX, 2022. 
This research was supported in part by the National Natural Science Foundation of China (Nos. 62072406, U21B2001, U19B2016), 
the Zhejiang Provincial Natural Science Foundation (No. LDQ23F020001), 
the National Key Laboratory of Science and Technology on Information System Security (No. 61421110502), 
the Key R\&D Programs of Zhejiang Province (No. 2022C01018), the National Key R\&D Projects of China (No. 2018AAA0100801).
\textit{(*Corresponding author: Shilian Zheng.)}

J.~Chen is with the Institute of Cyberspace Security, Zhejiang University of Technology, Hangzhou 310023, China (e-mail: chenjinyin@zjut.edu.cn).

J.~Ge is with the college of information engineering, Zhejiang University of Technology, Hangzhou 310023, China (e-mail: 2112103116@zjut.edu.cn).

S.~Zheng, W.~Shen, X.~Yang are with Innovation Studio of Academician Yang, National Key Laboratory of Electromagnetic Space Security, Jiaxing, China (e-mail: lianshizheng@126.com; beyondswg@163.com; yxn2117@126.com).

L.~Ye is with Binjiang Institute of Zhejiang University, Hangzhou 310023, China (e-mail: yelinhui1019@163.com).

H.~Zheng is with the Institute of Cyberspace Security, Zhejiang University of Technology, Hangzhou 310023, China (e-mail: haibinzheng320@gmail.com).

K.~Yue is with the Key Laboratory of RF Circuits and Systems, Ministry of Education, Hangzhou Dianzi University, Zhejiang, China (e-mail: kqyue@hdu.edu.cn).
}
}
\maketitle

\begin{abstract}
A wireless communications system usually consists of a transmitter which transmits the information and a receiver which recovers the original information from the received distorted signal. Deep learning (DL) has been used to improve the performance of the receiver in complicated channel environments and state-of-the-art (SOTA) performance has been achieved. However, its robustness has not been investigated. In order to evaluate the robustness of DL-based information recovery models under adversarial circumstances, we investigate adversarial attacks on the SOTA DL-based information recovery model, i.e., DeepReceiver. We formulate the problem as an optimization problem with power and peak-to-average power ratio (PAPR) constraints. We design different adversarial attack methods according to the adversary’s knowledge of DeepReceiver’s model and/or testing samples. Extensive experiments show that the DeepReceiver is vulnerable to the designed attack methods in all of the considered scenarios. Even in the scenario of both model and test sample restricted, the adversary can attack the DeepReceiver and increase its bit error rate (BER) above 10\%. It can also be found that the DeepReceiver is vulnerable to adversarial perturbations even with very low power and limited PAPR. These results suggest that defense measures should be taken to enhance the robustness of DeepReceiver.

\end{abstract}

% Note that keywords are not normally used for peerreview papers.

% For peer review papers, you can put extra information on the cover
% page as needed:
% \ifCLASSOPTIONpeerreview
% \begin{center} \bfseries EDICS Category: 3-BBND \end{center}
% \fi
%
% For peerreview papers, this IEEEtran command inserts a page break and
% creates the second title. It will be ignored for other modes.
\IEEEpeerreviewmaketitle

\section{Introduction}
%\subsection{Background and Motivation}
\IEEEPARstart{W}{ireless} communications plays an important role in both military and civil applications such as unmanned aerial vehicle control~\cite{khan2021role}, automatic driving~\cite{duan2020emerging}, positioning~\cite{gao2019mobile}, Internet of Things~\cite{nguyen2022Iot}, and cellular communications from the first generation (1G) to the fifth generation (5G)~\cite{ghosh20195G}. In a wireless communications system, information recovery plays an irreplaceable role. It refers to the process that the receiver recovers the original information bit stream sent by the transmitter from the received distorted signal which has been contaminated by noise, channel fading, and intentional/unintentional interference.

For traditional model-based information recovery methods, the receiver recovers the information from the received distorted signal through a series of processes, i.e., carrier and symbol synchronization~\cite{harris2000structure}, channel estimation~\cite{van1995channel}, equalization~\cite{chen1993clustering}, demodulation and channel decoding~\cite{hoeher1999turbo}. However, the recovering accuracy greatly depends on the performance of each processing module designed with theoretical assumptions which are difficult to be guaranteed in real-world scenarios. It is a challenge to precisely recover bit stream in real-world non-ideal scenarios.

Inspired by the success in various areas, deep learning (DL) has been considered as a promising candidate for information recovery to achieve better performance. DL-based information recovery methods are proposed to improve recovery performance in two fashions, non-end-to-end and end-to-end. Specifically, the non-end-to-end methods replace one or several modules of model-based information recovery with DL models to improve the performance of these modules. For example, 
DL-based channel estimation~\cite{bai2019deep, yang2019deep, mao2019roemnet,soltani2019deep, he2018deep, mehrabi2019decision, zamanipour2019survey}, channel equalization~\cite{zhang2020machine}, demodulation~\cite{fang2017deep, shental2019machine} and channel decoding~\cite{nachmani2016learning, nachmani2018near, nachmani2018deep, xu2018polar, doan2019neural} enhance the effect of corresponding modules, leading to improvement of recovery performance. The end-to-end methods, on the other hand, replace the entire information recovery modules with a single deep neural network (DNN). The recently proposed DeepReceiver~\cite{zheng2020deepreceiver} is one of such approaches which has achieved the state-of-the-art (SOTA) performance of information recovery in complex scenarios including radio frequency (RF) impairments, non-ideal noise, multi-path channel fading, and intentional co-channel interference.

Although DL-based information recovery has achieved competitive performance, its robustness under adversarial circumstances has not been investigated yet. In particular, an adversary generates malicious noises added to the original signal sent to the receiver to fool it in information recovery. These carefully crafted subtle perturbation added to the input signal is defined as an adversarial sample. In real-world applications, it may cause high bit error rate (BER) of information recovery and seriously affect the quality of the communication.

To investigate the issue of adversarial attacks on the DL-based information recovery, we propose various Adversarial attacks on SOTA Information Recovery model (AIR), i.e, DeepReceiver, in several scenarios, i.e., model knowledge unrestricted/restricted (ModelKnown/ModelUnknown) and testing samples knowledge unrestricted/restricted (TestKnown/TestUnknown). Furthermore, we consider the realistic realizability of the attacks, and propose power and peak-to-average power ratio (PAPR) restrictions to constrain the generated adversarial perturbations. We conduct extensive experiments to show the effects of these AIR methods on information recovery of DeepReceiver. The contributions of the paper are mainly as follows.

\begin{itemize}
\item To the best of our knowledge, this is the first work to study the robustness of the DL-based information recovery model, i.e., adversarial attacks on the SOTA information recovery model DeepReceiver, namely AIR. We formulate the problem of AIR as an optimization problem with the power and PAPR constraints. Different scenarios are considered to fully explore the vulnerability of the information recovery model in the face of AIR.
%propose four attack methods according to the practical scenarios to fully explore the vulnerability of the recovery model.
%the attack on the information recovery model is divided into four scenarios and the corresponding attack methods are designed for each scenario.

\item We design different AIR methods according to the adversary's knowledge of DeepReceiver's model and/or testing samples. In the scenario of ModelKnown and TestKnown, we design three multi-loss white-box attack methods according to the structure of DeepReceiver. In the scenario of MoldelUnknown and TestKnown, we use a surrogate model to replace the target model for attack. In the scenario of TestUnknown, we use a universal adversarial perturbation (UAP) method with only training samples to attack DeepReceiver at inference stage. In these designed methods, we propose to clip the maximum amplitude of the adversarial perturbations to restrict the PAPR and use power normalization to restrict the power of the adversarial perturbations.
% \item Considering the physical feasibility of adversarial perturbations transmission in transmitter, we use the power normalization method to restrict the adversarial perturbation's power, and restrict the PAPR through restrict the maximum amplitude. Additionally, ......

\item Extensive experiments are conducted to verify the AIR's effects on DeepReceiver. Results show that the DeepReceiver model is vulnerable to AIR in all of the considered scenarios. Even in the scenario of TestUnknown (and ModelUnknown), the adversary can attack the DeepReceiver and increase its BER above 10$\%$. It can also be found that the DeepReceiver is vulnerable to adversarial attacks even with very low power and limited PAPR. These results suggest that defense measures should be taken to enhance the robustness of DeepReceiver.
\end{itemize}

The remainder of this paper is organized as follows. We describe the related work in Section~\ref{Related Work}. Section~\ref{system model and attack scenario} briefly introduces the DeepReceiver model and the formulated problem. In Section~\ref{method}, we describe the designed AIR methods in detail. Experimental results are provided in Section~\ref{experimental} and conclusions are made in Section~\ref{conclusion}.

\section{Related Work \label{Related Work}}
DNNs have been proved vulnerable to adversarial attacks~\cite{szegedy2013intriguing}, that is, they are easily fooled by carefully crafted adversarial samples. 
The research of adversarial attack on DL-based wireless communications system is gradually expanding which mainly focused on the DL-based signal modulation cclassifiers~\cite{kim2020over,lin2020threats,hameed2020best,flowers2019evaluating,kokalj2019adversarial,usama2019adversarial,kim2020channel,kim2020adversarial,sadeghi2018adversarial} and DL-based spectrum sensing models~\cite{shi2018spectrum,sagduyu2019iot,sagduyu2019adversarial,Haghighat2015,zheng2021primary}.

The attack on DL-based modulation classifiers uses the constructed adversarial signal to make the classifiers incorrectly recognize the modulation type of the input signal, such as recognizing the input BPSK signal as QPSK. 
Numerous white-box attack methods are proposed to deceive the DL-based signal modulation classifiers. Specifically, Kim~\emph{et al.}~\cite{kim2020over} used fast gradient method (FGM) to attack the modulation classifiers.  Both targeted and untargeted attacks were considered. Targeted attack has been used by enforcing the DNNs to misclassify the input sample to a target label.
Sadeghi~\emph{et al.}~\cite{sadeghi2018adversarial} used the fast gradient sign method (FGSM) attack, which is the earliest adversarial attack method proposed on deep learning to attack modulation classifiers.
Lin~\emph{et al.}~\cite{lin2020threats} used the basic iterative method (BIM) attack to verify the effectiveness of adversarial attacks on modulation classifiers and modulation datasets. Hameed~\emph{et al.}~\cite{hameed2020best} used the project gradient descent (PGD) attack to generate adversarial samples, which is one of the strongest first-order attack methods. Besides, some black-box attack methods, such as UAP~\cite{kim2020over} and PCA-based UAP~\cite{sadeghi2018adversarial}, are also successfully conducted on DL-based signal modulation classifiers. For example,
Kim~\emph{et al.}~\cite{kim2020over} and Sadeghi~\emph{et al.}~\cite{sadeghi2018adversarial} used universal adversarial perturbations (UAP) and PCA-based UAP, respectively, and designed the input-independent adversarial perturbations and showed the vulnerability of modulation classifiers.

For DL-based spectrum sensing models, there are two types of attacks, spectrum poisoning attack~\cite{shi2018spectrum,sagduyu2019iot,sagduyu2019adversarial,Haghighat2015} and spectrum adversarial attack~\cite{zheng2021primary}. The spectrum poisoning attack is realized by modifying the training data of the spectrum sensing model, that is, the adversary leaves a backdoor in the spectrum sensing model which was trained based on the poisoned training data, and makes the spectrum sensing model output the wrong result when fed by some specific triggered samples. 
Spectrum adversarial attack refers to the adversary adds carefully manufactured subtle disturbances to the benign primary user signal, which makes the secondary user misdetect the primary user signal and access the channel to cause harmful interference to the primary user. 
Zheng~\emph{et al.}~\cite{zheng2021primary} proposed three black-box attack methods on spectrum sensing model, showed that even if the decision boundary of the model is changed, adversarial samples can still attack spectrum sensing model successfully.
%The difference between spectrum poisoning attack and spectrum adversarial attack is that spectrum poisoning attack is aimed at the training stage, while spectrum adversarial attack is aimed at the inference stage.

In summary, the existing attack methods on DL-based wireless communications system are mainly focused on adversarial attacks and poisoning attacks on either DL-based modulation classifiers or DL-based spectrum sensing models, which can not be directly applied to the information recovery model which is the focus of this paper. The vulnerability of DL-based information recovery models such as DeepReceiver in the face of adversarial attacks is still an open problem.
%From the above, it can be found that the researches on attack in the field of WCS is mainly focus on signal modulation classification task and spectrum sensing task. Different from these researches, our work focuses on the information recovery.

%\section{Preliminaries\label{system model and attack scenario}}
\section{System Model\label{system model and attack scenario}}
\subsection{DeepReceiver\label{deepreceiver}}
 Wireless communications system generally consists of a transmitter and a receiver. At the transmitter, the content such as text, audio or video that needs to be sent is converted into the information bit stream $s = \left[ s _ { 1 } , s _ { 2 } \ldots, s _ { M } \right]$ by source coding and encryption, where $M$ is the number of bits in the stream and $s _ { m } \in \Theta = \{ 0,1 \}$. After channel encoding, modulation and pulsed shaping, the information bit stream is transformed into a signal which is transmitted to the receiver through the wireless channel. The received sampled baseband signal can be represented as
 \begin{equation}
r(n) = x(n)*h_1(n)e^{j(2\pi\Delta f n+\theta)} +w(n),
\end{equation}
 
 \noindent where $x(n)$ is the transmitted baseband signal, $h_1(n)$ is the impulse response of the channel between the transmitter and the receiver, $\Delta f$ is the frequency offset, $\theta$ is the phase deviation, $w(n)$ is additive white Gaussian noise (AWGN), and $*$ represents convolution operation.
 
 DeepReceiver is a receiver model based on DNN for information recovery of the signal generated by the transmitter. The in-phase (I) component and quadrature (Q) component of the received signal are decomposed as
 \begin{equation}
 I_r(n)=\mathrm{Re}(r(n)), 
 Q_r(n)=\mathrm{Im}(r(n)),
 \end{equation}
 where $\mathrm{Re}(r(n))$ represents the real part of the received signal and $\mathrm{Im}(r(n))$ represents the imaginary part of the received signal. The IQ components are then fed into the DeepReceiver to output the recovered information bit stream $ \hat { { s } } = \left[ \hat { s } _ { 1 } , \hat { s } _ { 2 } , \ldots , \hat { s } _ { M } \right] , \hat { s } _ { m } \in \Theta $. With the goal of minimizing bit error rate (BER), the optimization of DeepReceiver can be expressed as
%The DeepReceiver structure is shown in Fig.~\ref{fig2a}. The digital sampled IQ signal is used as the input of the CNN. After a series of operations of convolution, pooling, and activation, a feature vector is obtained. The feature vector is used as the input of the M binary classifiers, and the outputs of the binary classifiers correspond to the recovered information bit stream. 
% At the transmitter, the content such as text, audio or video that needs to be sent is converted into information bit stream which can be denoted as $s = \left[ s _ { 1 } , s _ { 2 } \ldots s _ { M } \right]$ after source coding and encryption, where $M$ is the number of bits in the stream and $s _ { m } \in \Theta = \{ 0,1 \}$. After channel encoding, modulation and pulsed shaping, the information bit stream is transformed into signal and then radiated into the air via the antenna. The signal is transmitted to the receiver through the wireless channel. The receiver inputs the received IQ signal into the DeepReceiver, and outputs the recovered information bit stream $ \hat { { s } } = \left[ \hat { s } _ { 1 } , \hat { s } _ { 2 } , \ldots , \hat { s } _ { M } \right] , \hat { s } _ { m } \in \Theta $. With the goal of minimizing BER, the training process of DeepReceiver can be expressed as follow:
\begin{equation}
\min _ { \mathcal{Q} } \| \hat {{ s } } - { s } \| _ { 1 } , \hat {{ s } } = \mathcal{F} ({ r(n) }; \mathcal { Q }),
\end{equation}

\noindent where $\mathcal{Q}$ represents the parameters of DeepReceiver, $\| \bullet \|_1$ represents the operation of norm 1, and $\mathcal{F}$ represents the function of DeepReceiver from input to output. Note that the last part of DeepReceiver adopts $M$ binary classifiers to recovery $M$-bit information. The sum of the cross-entropy of the $M$ classifiers is used as the loss function to train DeepReceiver, i.e.,
\begin{equation}
\mathrm{ loss } \left( \mathcal{R}(n), \mathcal{D} \right) = - \frac { 1 } { N_B } \sum _ { i = 1 } ^ { N_B } \sum _ { m = 1 } ^ { M } \sum _ { k = 1 } ^ { 2 } d _ { i m k } \log \left( c _ { i m k } \right), \label{formal 2}
\end{equation}

\noindent where $\mathcal{R}(n) = \left \{r_i ( n )|i=1,2,\ldots,N_B\right \}$ represents a mini-batch of $N_B$ samples, $\mathcal{D}=\{d _ { i m k }| i = 1,2,\ldots,N_B; m = 1,2,\ldots,M; k = 1, 2 \}$ represents the labels of the mini-batch of samples, $d _ { i m k }$ is the $k$-th true label corresponding to the $m$-th bit of the $i$-th sample, and $c _ { i m k }$ represents the output probability of the $m$-th classifier on the $k$-th class when the $i$-th sample is used as the input.

\subsection{Problem Formulation\label{attack scenario}}
\begin{figure*}[htbp]
    \centering
    \includegraphics[width=0.8\textwidth]{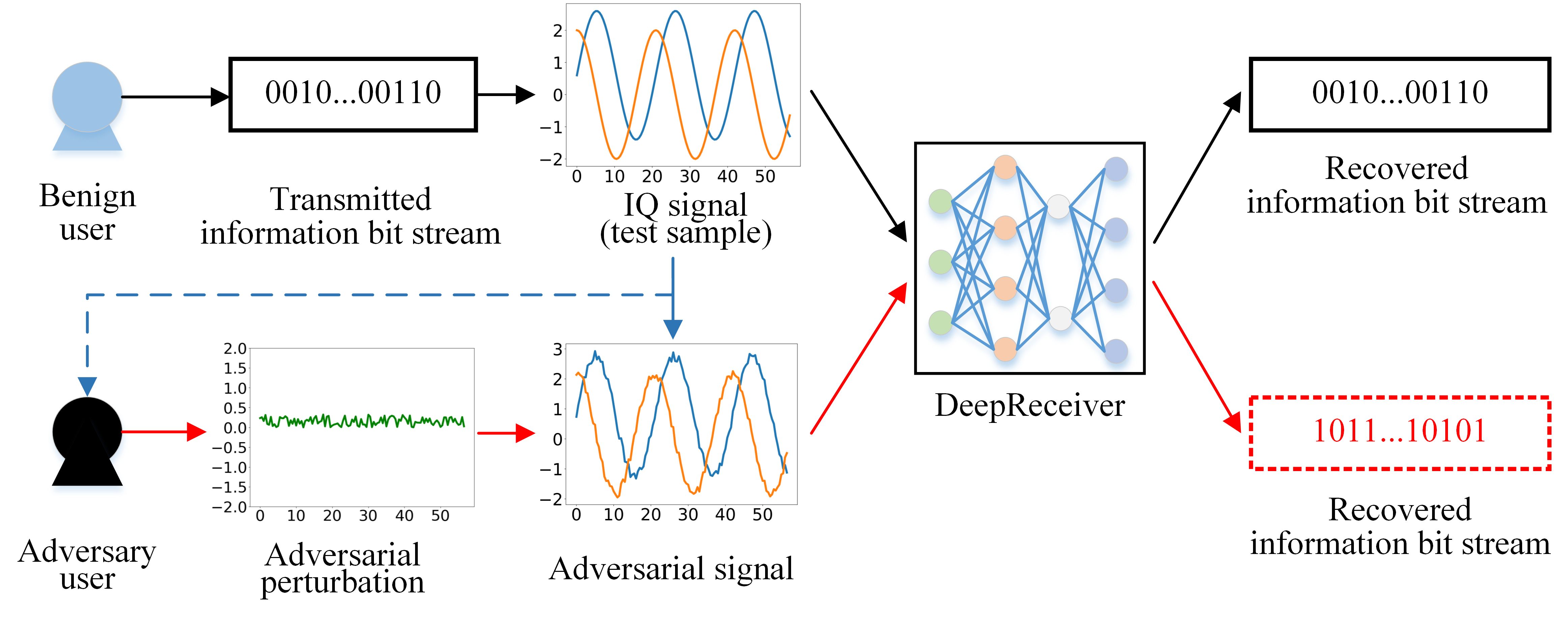}
    \caption{Attack framework on DeepReceiver.}
    \label{attack_scene}
\end{figure*}

When attacking DeepReceiver, the adversary sends a constructed adversarial perturbation through the adversary's transmitter which is superimposed on the signal sent by the benign user to become an adversarial signal for DeepReceiver. The concept of adversarial attack on DeepReceiver is shown in Fig.~\ref{attack_scene}. The general purpose of the attack is to make DeepReceiver output the wrong information bit stream and maximize the BER. The problem can be formulated as follows:
\begin{equation}
\max _ { \rho  ( n ) } \| \mathcal{F} ( r ( n ) + h_2(n)*\rho  ( n ) ) - \mathcal{F} ( r ( n ) ) \| _ { 1 }, \label{formal 5}
\end{equation}

\noindent where $\rho ( n )$ represents the adversarial perturbation launched by the attacker, $h_2(n)$ is the channel response between the adversary's transmitter and DeepReceiver. With an ideal AWGN channel assumption, $h_2(n)$ is simply a constant, denoted as $\gamma$, which represents the channel gain. In this case, the perturbation arrived at the DeepReceiver side can be represented as
\begin{equation}
\delta (n)= \gamma \rho (n). 
\label{formal 5.1}
\end{equation}

The adversary needs to send the adversarial perturbation through the transmitter. In the real-world attack scenario, due to the limitations of the transmitter's hardware conditions such as nonlinearity of the power amplifier (PA), the transmitter cannot transmit signals with excessive power and PAPR with high fidelity. Therefore, it is necessary to restrict the power and PAPR of the adversarial perturbation. The power of adversarial perturbation is defined as
\begin{equation}
\mathrm{Power} ( \rho ( n ) ) = \frac { \sum _ { n = 1 } ^ { N }  |\rho ( n )| ^ { 2 } } { N },
\end{equation}

\noindent where $N$ represents the length of adversarial perturbation which is the same as $r(n)$. The definition of PAPR is
\begin{equation}
\mathrm { PAPR } ( \rho ( n ) ) = \frac { \max \left( |\rho ( n )| ^ { 2 } \right) } { \sum _ { n = 1 } ^ { N } |\rho ( n )| ^ { 2 } / N }. \label{formal 4}
\end{equation}

According to (\ref{formal 5.1}), restricting the power and PAPR of $\rho (n)$ equivalents to restricting the power and PAPR of $\delta (n)$, thus the adversarial attack on DeepReceiver can be expressed as follows:
\begin{equation}
\begin{array} { l } \delta ( n )=\underset{ \delta  ( n ) }{\arg \max}  \| \mathcal{F} ( r ( n ) + \delta (n)) - \mathcal{F} ( r ( n ) ) \| _ { 1 }, \\ 
\text { s.t. } \quad \operatorname {Power} ( \delta ( n ) ) \le \varepsilon, \\ 
\quad \quad \quad \operatorname {PAPR} ( \delta ( n ) ) \le \beta ,
\end{array} 
\label{formal 9}
\end{equation}
\noindent where $\varepsilon$ and $\beta$ represent the power limitation and PAPR limitation of adversarial perturbation respectively.

\section{Adversarial Attacks on DeepReceiver\label{method}}
\subsection{Overview of AIR\label{attack framework}}
In order to systematically explore the robustness of DeepReceiver, according to whether the adversary can obtain the prior knowledge of the model and/or the testing samples, we divide the adversarial attacks into the following four scenarios: ModelKnown, ModelUnknown, TestKnown and TestUnknown.

%\textbf{Model knowledge unrestricted}.
\begin{itemize}
\item \textbf{ModelKnown}.
The adversary can obtain the DeepReceiver model's structure and parameters $\mathcal {Q}$. If the adversary also has the knowledge of the testing samples, white-box attacks can be used.

%\textbf{Model knowledge restricted}. 
\item \textbf{ModelUnknown}.
The adversary can't obtain the DeepReceiver model's structure parameters $\mathcal {Q}$. In general, it is difficult for an adversary to obtain the complete structure and parameters of the target DeepReceiver model. However, a surrogate model may be constructed which can be used to attack.

%\textbf{Test samples knowledge unrestricted}. 
\item \textbf{TestKnown}.
The adversary can obtain the signal received at the DeepReceiver $r(n)$. The adversary can feed the testing samples into the DeepReceiver model or the constructed surrogate model to generate adversarial perturbations via gradient-based methods.

%\textbf{Test samples knowledge restricted}. 
\item \textbf{TestUnknown}.
The adversary can't get the signal received at the DeepReceiver $r(n)$. 
In this case, the adversary needs to generate generalized adversarial perturbation which can attack all possible testing samples.
\end{itemize}

In each attack scenario, different AIR methods are designed to attack DeepReceiver. Specifically, when the adversary is able to obtain the testing samples, DeepReceiver model's structure and parameters (TestKnown and ModelKnown), three gradient-based attack methods are adopted to implement adversarial attack on DeepReceiver. The FGSM~\cite{goodfellow2014explaining} attack method has the advantage of fast convergence speed, but it has the problem of weak attack ability. Therefore, momentum iterative fast gradient sign method (MI-FGSM)~\cite{dong2018boosting} and PGD~\cite{madry2017towards} attack method with stronger attack ability are further conducted to explore the robustness of DeepReceiver. When the testing sample knowledge is unrestricted but the model knowledge is restricted (TestKnown and ModelUnknown), the attack is carried out by constructing a surrogate model of DeepReceiver. The above-mentioned three gradient-based methods are used to attack the constructed surrogate model to generate adversarial perturbations which are then used to attack the DeepReceiver through the attack transferability of adversarial perturbations. Finally, since UAP attack method~\cite{moosavi2017universal} does not depend on the testing samples, the DeepReceiver is attacked through UAP when the knowledge of the testing sample is restricted (TestUnknown).
The specific attack methods used in each scenario are summarized in Table~\ref{table1}.

% Please add the following required packages to your document preamble:
\begin{table}[htbp]
    %\normalsize
    \centering
    \caption{Overview of Attack Methods}
    \renewcommand{\arraystretch}{1.2}
    \begin{tabular}{m{1.1cm}<{\centering} m{1.7cm}<{\centering} m{1cm}<{\centering} m{3cm}<{\centering}}
    %\begin{tabular}
    \hline\cline{1-4}
    \textbf{Scenario}   & \textbf{Testing Sample}   & \textbf{Model}          & \textbf{Attack Method} \\
    \hline\cline{1-4}
    ~                   & ~                         & ~                       & FGSM~\cite{goodfellow2014explaining} \\
    A                   & \Large\checkmark          & \Large\checkmark        & MI-FGSM~\cite{dong2018boosting} \\
    ~                   & ~                         & ~                       & PGD~\cite{madry2017towards} \\
    \hline\rule{0pt}{10pt}
    ~                   & ~                         & ~                       & Surrogate+FGSM \\ 
    B                   & \Large\checkmark          & \Large\XSolidBrush      & Surrogate+MI-FGSM \\
    ~                   & ~                         & ~                       & Surrogate+PGD \\ 
    \hline\rule{0pt}{10pt}
    C                   & \Large\XSolidBrush        & \large$/$               & UAP~\cite{moosavi2017universal} \\
    \hline\cline{1-4}
    \end{tabular}
    \label{table1}
\end{table}

It's important to note that in attacking DeepReceiver, additional processing needs to be considered. The first is to deal with multiple binary classifiers in DeepReceiver model rather than single classifier that FGSM, MI-FGSM, PGD and UAP are originally designed for. The second is to deal with the power and PAPR constraints of the generated adversarial perturbations, shown in Eq. (\ref{formal 9}). Details of these will be discussed in the next subsections.

\subsection{AIR in TestKnown and ModelKnown Scenario}
%\subsubsection{ModelKnown}
In the TestKnown and ModelKnown scenario, the adversary can quickly construct the adversarial perturbation by obtaining the gradient of the DeepReceiver model's loss function to the input. We explore the attack effect of FGSM~\cite{goodfellow2014explaining}, MI-FGSM~\cite{dong2018boosting}, and PGD~\cite{madry2017towards} on DeepReceiver. The reason for using these attack methods in this scenario is that on the one hand, they are representative in white-box attack methods, and on the other hand, they have relatively high attack efficiency.

\subsubsection{FGSM-based attack method}

FGSM was first proposed by Goodfellow \emph{et al.}~\cite{goodfellow2014explaining} to generate adversarial samples. The gradient is obtained by calculating the derivative of the model's loss function to the input, and then the specific gradient direction is obtained by the sign function, and finally the gradient is multiplied by the step size to obtain the adversarial perturbation. The FGSM attack method can be expressed as follows:
\begin{equation}
x _ { adv } = x + \partial \cdot \operatorname { sign } \left( \nabla _ { x } J ( \theta , x , y ) \right),
\end{equation}

\noindent where $\partial$ represents the step size, $\operatorname { sign } ( \cdot )$ represents sign function, $J ( \theta , x , y )$ represents the loss function of the model, $\theta$ represents the parameters of the model, $x$ is the input, $y$ is the true label, and $\nabla _ { x }$ represents the derivative of the loss function $J ( \theta , x , y )$ to the input $x$.

FGSM attack method constructs adversarial perturbation through gradient, and is originally designed for attacking models with a single input and a single output. Differently, the last layer of DeepReceiver is composed of $M$ classifiers to recover $M$ bits information stream. In other words, there is only one input of DeepReceiver, but $M$ outputs. Nevertheless, according to the FGSM attack method, adding noise to the input in the direction of gradient can make the value of loss function rise rapidly and make the model output the wrong result. Thus, the adversarial perturbation to DeepReceiver can be constructed by obtaining the gradient of the input to the loss function shown in the following:
\begin{equation}
\operatorname { loss} _ { a d v } \left( r ( n ) , d _ { m k } \right) = - \sum _ { m = 1 } ^ { M } \sum _ { k = 1 } ^ { 2 } d _ { m k } \log \left( c _ { m k } \right),
\end{equation}

\noindent where $c _ { m k }$ is the output probability of the $m$-th classifier on the $k$-th category, $d _ { m k }$ represents the $k$-th true label corresponding to the $m$-th bit of the input signal $r ( n )$. The constructed adversarial perturbation is
\begin{equation}
\hat {\delta} ( n ) = \operatorname { sign } \left( \nabla _ { r ( n ) } \operatorname { loss } _ { a d v } \left( r ( n ) , d _ { m k } \right) \right).
\label{ap1}
\end{equation}

After the adversarial perturbation is obtained, its PAPR and power need to be restricted. It should be noted that the PAPR of the adversarial perturbation generated by Eq. (\ref{ap1}) is 1, which is the possible minimum value, so there is no need to restrict PAPR. For the power restriction of the adversarial perturbation, it can be achieved through power normalization which can be represented as
\begin{equation}
\widetilde {\delta} ( n ) = \frac { \hat {\delta} ( n ) } { \sqrt { \operatorname {Power} \left (\hat {\delta} ( n ) \right ) }}.
\label{e4fgsm}
\end{equation}

\noindent If we want to add a specified power of adversarial perturbation to the testing sample, we only need to multiply $\widetilde {\delta} ( n )$ by $\sqrt { \varepsilon }$ and add it to the testing sample. Therefore, the final adversarial perturbation after power restriction is 
\begin{equation}
{\delta} ( n ) = \sqrt { \varepsilon } \cdot \widetilde {\delta} ( n ),
\label{e5fgsm}
\end{equation}
and the final adversarial signal is
\begin{equation}
r _ { a d v } ( n ) = r ( n ) + {\delta} ( n ).
\end{equation}

\subsubsection{MI-FGSM-based attack method}
%FGSM is characterized by fast attack speed, because FGSM does not require iteration and only needs to calculate the gradient once to construct adversarial perturbations. However, the constructed adversarial signal has the problem of weak attack ability. In TestKnown and ModelKnown scenario, we further explore the attack effect of MI-FGSM and PGD on DeepReceiver, which are two attack methods with stronger attack ability.

MI-FGSM integrates momentum into FGSM attack method. In each iteration, the velocity vector is accumulated in gradient direction, which can make the gradient rise rapidly and generate a more robust adversarial perturbation. MI-FGSM updates adversarial perturbation $\delta ( n )$ iteratively as
\begin{equation}
\delta_{ t + 1 } ( n ) = \operatorname { sign } \left( g _ { t + 1 } \right) + \delta_ { t }( n ),
\label{e1mifgsm}
\end{equation}
\begin{equation}
r_{adv,t+1}(n)=r_{adv,t}(n)+ \operatorname{PowerRes}(\delta _{t+1}(n)),
\label{e2mifgsm}
\end{equation}
where $t$ is the current number of iterations,
$\operatorname{PowerRes}(\cdot)$ represents operations for PAPR and Power restrictions and $g _ {t+1}$ is
\begin{equation}
g _ { t + 1 } = g _ { t } + \frac { \nabla _ { r _ {adv,t}(n) } \operatorname { loss } _ { a d v } \left( r _ {adv,t}(n), d _ { m k } \right) } { \left\| \nabla _ { r _ {adv,t}(n) } \operatorname { loss } _ { a d v } \left( r _ {adv,t}(n), d _ { m k } \right) \right\| _ { 1 } }.
\label{e3mifgsm}
\end{equation}

As for PAPR restriction, as long as the maximum amplitude of the adversarial perturbation is restricted, the PAPR of the adversarial perturbation can be restricted. In other words, if the PAPR of the signal is restricted to $\beta$, the maximum amplitude of the signal needs to satisfy
\begin{equation}
% \operatorname{A_{max}} ( \hat{\delta} ( n ) , \beta ) = \sqrt {{ \beta } \cdot \sum _ { n = 1 } ^ { N } |\hat{\delta} (n)| ^ { 2 } / N }, 
\operatorname{A_{max}} ( \hat {\delta} ( n ) , \beta ) = \sqrt {{ \beta } \cdot \operatorname{Power}( \hat {\delta} ( n ) )}. 
\label{e4mifgsm}
\end{equation}
\noindent Therefore, we use the following operation to restrict the PAPR of the adversarial perturbation:
\begin{equation}
\bar {\delta} ( n ) = \operatorname { clip } ( \hat {\delta} ( n ) , \operatorname{A_{max}} ( \hat {\delta} ( n ) , \beta ) ),
\label{e5mifgsm}
\end{equation}
\noindent where $\operatorname { clip } ( x , y )$ means restricting the amplitude of $x$ no larger than $y$. As for power restriction, the process is the same as that discussed in FGSM. In summary, the process of using MI-FGSM to generate adversarial perturbations is shown in Algorithm ~\ref{algorithm1}.

\begin{algorithm}[htbp]
\caption{MI-FGSM attack method}
\begin{algorithmic}[1]

\renewcommand{\algorithmicrequire}{\textbf{Input:}}
\REQUIRE Testing samples $r(n)$, loss function $\operatorname{loss}_{adv}$, true label $d_{mk}$, and iteration number $T$.

\renewcommand{\algorithmicrequire}{\textbf{Restrictions:}}
\REQUIRE Power restriction $\varepsilon$ and PAPR restriction $\beta$.

\renewcommand{\algorithmicensure}{\textbf{Output:}}
\ENSURE Adversarial signal $r_{adv}(n)$.

\STATE $g_{0}=0$, $r_ { adv,0 }( n ) = r ( n )$, $\delta_ { 0 } ( n )  = 0$.

\FOR{$t=0$ to $T-1$}
    \STATE Input $r_ { adv,t }( n )$ to DeepReceiver and obtain the gradient $\nabla _ { r_ { adv,t }( n ) } \operatorname { loss } _ { a d v } \left( r_ { adv,t }( n ) , d _ { m k } \right)$.

    \STATE Update $g_{t+1}$ by accumulating the velocity vector in gradient by Eq.~(\ref{e3mifgsm}).
    %$g _ { t + 1 } = g _ { t } + \frac { \nabla _ { r_ { adv,t }( n ) } \operatorname { loss } _ { a d v } \left( r_ { adv,t }( n ) , d _ { m k } \right) } { \left\| \nabla _ { r_ { adv,t }( n )} \operatorname { loss } _ { a d v } \left( r_ { adv,t }( n ) , d _ { m k } \right) \right\| _ { 1 } }$

    \STATE Update adversarial perturbation $\delta_{ t + 1 } ( n )$ by Eq.~(\ref{e1mifgsm}).
    % $\delta_{ t + 1 } ( n ) = \operatorname { sign } \left( g _ { t + 1 } \right) + \delta_ { t }( n )$
    
    \STATE Restrict the PAPR of the adversarial perturbation by Eq.~(\ref{e5mifgsm}).

    \STATE Normalize the power of the adversarial perturbation by Eq.~(\ref{e4fgsm}).
    
    \STATE Update $r_ { adv,t+1 }( n )$ by Eq.~(\ref{e2mifgsm}).
    
        %$r_ { adv,t+1 }( n ) = r_ { adv,t }( n ) + \sqrt { \varepsilon ^ { \prime } } \cdot \delta _ { t + 1 } ^ { \prime \prime } ( n )$
\ENDFOR

\RETURN $r _ { a d v } ( n ) = r _ { adv, T }( n )$
\end{algorithmic}
\label{algorithm1}
\end{algorithm}

\subsubsection{PGD-based attack method}
PGD divides the process of calculating gradient once in FGSM-based attack method into several times to construct adversarial perturbation. In other words, the next adversarial signal is based on the gradient update of the previous adversarial signal. PGD updates adversarial perturbation $\delta ( n )$ iteratively as
\begin{equation}
\delta_{ t + 1 } ( n ) = \operatorname{sign} \left( g \right) + \delta_ { t }( n ),
\label{e1pgd}
\end{equation}
\begin{equation}
r_{adv,t+1}(n)=r_{adv,t}(n)+ \operatorname{PowerRes}(\delta _{t+1}(n)),
\label{e2pgd}
\end{equation}
where $g$ is 
\begin{equation}
g = \nabla _ { r _ {adv, t } (n)} \operatorname { loss } _ { a d v } \left( r _ { adv,t }(n), d _ { m k } \right).
\label{e3pgd}
\end{equation}

As for the PAPR and power restriction, the process is the same as that discussed in MI-FGSM and FGSM, respectively. The process of using PGD attack methods to generate adversarial perturbation is shown in Algorithm ~\ref{algorithm2}.

% FGSM is characterized by fast attack speed, because FGSM does not require iteration and only needs to calculate the gradient once to construct adversarial perturbations. However, the constructed adversarial signal has the problem of weak attack ability. Under the condition that the test samples knowledge and model knowledge are unrestricted, we further explored the attack effect of MI-FGSM and PGD on DeepReceiver, which are two attack methods with stronger attack ability. Similarly, we use MI-FGSM and PGD attack methods to construct the adversarial perturbations through the loss function of DeepReceiver. The process of using MI-FGSM and PGD attack methods to generate adversarial perturbations is shown in Algorithm ~\ref{algorithm1} and Algorithm ~\ref{algorithm2}, respectively. MI-FGSM integrates momentum into FGSM attack method. In each iteration, velocity vector is accumulated in gradient direction, which can make the gradient rise rapidly and generate a more robust adversarial perturbations. PGD divides the process of calculating gradient once to construct adversarial perturbations in FGSM attack method into several times of calculating gradient to construct adversarial perturbations. In other words, the next adversarial signal is based on the gradient update of the previous adversarial signal. Therefore, PGD can generate a more refined adversarial signal than FGSM.

\begin{algorithm}[htbp]
\caption{PGD attack method}
\begin{algorithmic}[1]
\renewcommand{\algorithmicrequire}{\textbf{Input:}}
\REQUIRE Testing samples $r(n)$, loss function $\operatorname{loss}_{adv}$, true label $d_{mk}$, and iteration number $T$.

\renewcommand{\algorithmicrequire}{\textbf{Restrictions:}}
\REQUIRE Power restriction $\varepsilon$ and PAPR restriction $\beta$.

\renewcommand{\algorithmicensure}{\textbf{Output:}}
\ENSURE Adversarial signal $r_{adv}(n)$.

\STATE $g_{0}=0$, $r _ { adv,0 }( n ) = r ( n )$, $\eta = \varepsilon / T$, $\delta _ { 0 }( n ) = 0$.

\FOR{$t=0$ to $T-1$}
    \STATE Input $r _ { adv,t }( n )$ to DeepReceiver and obtain the gradient $g$ by Eq.~(\ref{e3pgd}).

    % \STATE Update $g$ by Eq.~(\ref{e3pgd}).
    
    \STATE Update adversarial perturbation $\delta_{ t + 1 } ( n )$ by Eq.~(\ref{e1pgd}).
    
    \STATE Restrict the PAPR of the adversarial perturbation by Eq.~(\ref{e5mifgsm}).
    
    \STATE Normalize the power of the adversarial perturbation by Eq.~(\ref{e4fgsm}).
    
    \STATE Update $r_ { adv,t+1 }( n )$ by Eq.~(\ref{e2pgd}).
    
    % \STATE Obtain adversarial perturbation $\delta_{ t + 1 } ( n ) = \operatorname { sign } \left( g_{t+1} \right) + \delta_ { t }( n )$ under the current iteration number and obtain $\delta_{ t + 1 } ^ { \prime \prime } ( n )$ after restricting the PAPR and power of $\delta_ { t + 1 }( n )$.

    % \STATE Update $s _ { t + 1 } ^ { * } ( n )$ as:

    %     $s _ { t + 1 } ^ { * } ( n ) = s _ { t } ^ { * } ( n ) + \sqrt { \varepsilon ^ { \prime } } \cdot \delta _ { t + 1 } ^ { \prime \prime } ( n )$

\ENDFOR

\RETURN $r _ { a d v } ( n ) = r _ { adv,T } ( n )$
\end{algorithmic}
\label{algorithm2}
\end{algorithm}

\begin{figure*}[htbp]
  \centering
  \subfigure[]{
  \includegraphics[height=0.20\textwidth]{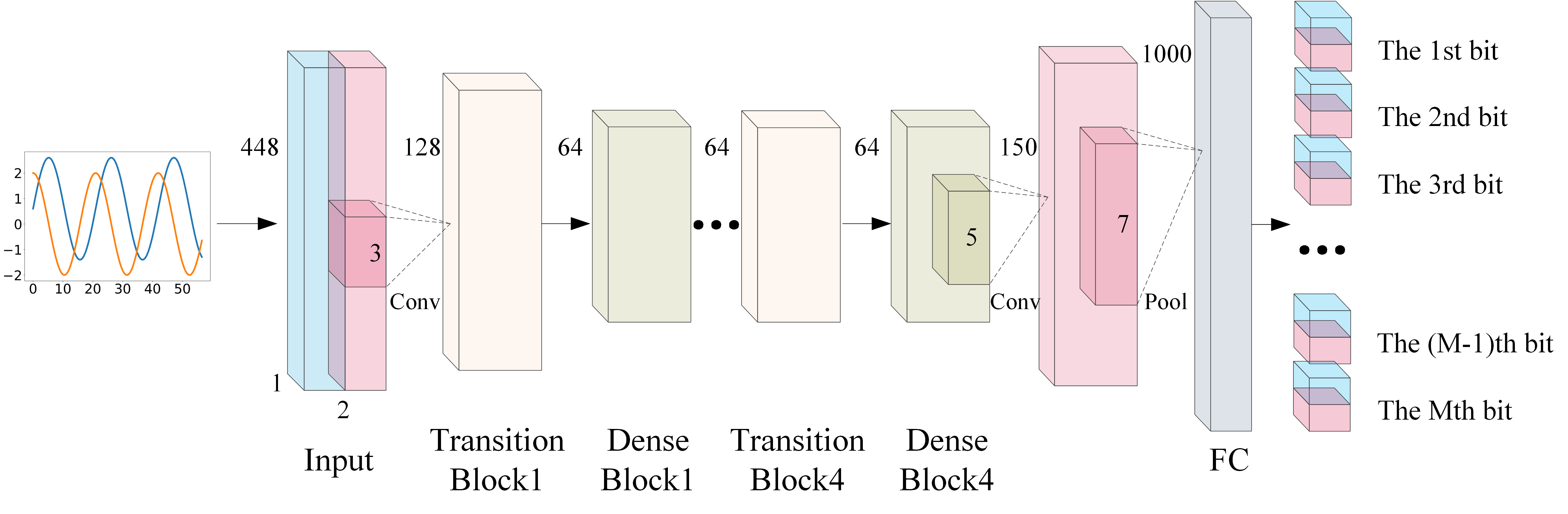}
  \label{fig2a}
    }
  \subfigure[]{
  \includegraphics[height=0.25\textwidth]{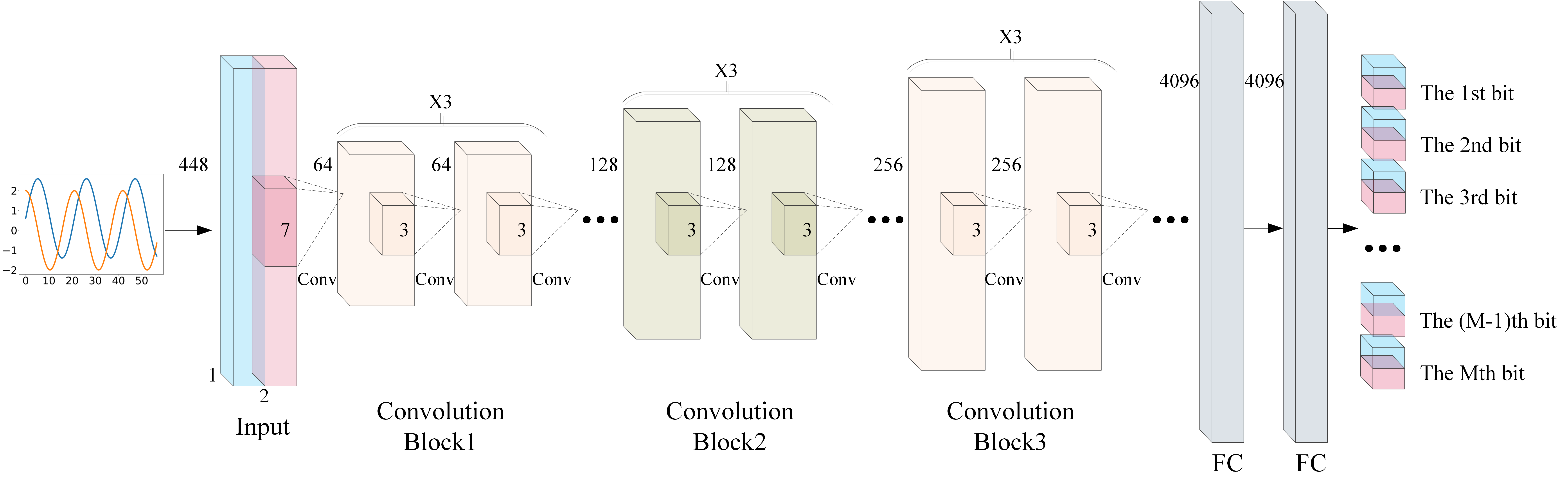}
  \label{fig2b}
    }
  \caption{Comparison of network structure between DeepReceiver and the surrogate model. (a) Model structure of DeepReceiver and (b) Model structure of the surrogate model.}
\end{figure*}

\subsection{AIR in TestKnown and ModelUnknown Scenario}

When the testing samples knowledge is unrestricted but the model knowledge is restricted, the adversarial can attack the DeepReceiver through constructing a surrogate model. Specifically, the DeepReceiver is attacked by the attack transferability of the adversarial signal generated on the surrogate model of DeepReceiver. The surrogate model refers to a model with the same function as the DeepReceiver, which means that the surrogate model constructed by the adversary can also recover the original information bit stream from the received signal. The network structure of the DeepReceiver is shown in Fig.~\ref{fig2a}. Its input is an IQ signal, and the output layer completes the recovery of the information bit stream through $M$ binary classifiers. The main structure of DeepReceiver consists of several transition blocks and dense blocks. Under the condition of ModelUnknown, in the surrogate model, we can not construct a similar structure as DeepReceiver. Instead, we replace the main part of the DeepReceiver with other structures, such as ResNet~\cite{he2016deep}. An example of a surrogate model is shown in Fig.~\ref{fig2b}. The input is the received IQ signal and the output layer also adopts $M$ binary classifiers for the recovery of the information bit stream. However, the main structure of the surrogate model constructed is almost completely different from that of the DeepReceiver. The training process of the surrogate model is the same as the DeepReceiver. The surrogate model is model knowledge unrestricted for the adversary. Therefore, after the training is completed, we use FGSM, MI-FGSM and PGD attack methods to attack the surrogate model to obtain adversarial perturbations which are then used to attack the DeepReceiver.

\subsection{AIR in TestUnknown Scenario}
In the process of attacking DeepReceiver, the adversary may not be able to obtain testing samples, so the aforementioned gradient-based attack methods with or without surrogate model can not be used to construct the adversarial signals. In this case, we use UAP \cite{moosavi2017universal} to attack the DeepReceiver because UAP can attack the model with only training samples. It should be noted that because the UAP attack method can attack the model regardless of whether the model knowledge is restricted, when using UAP to attack, we no longer consider whether the model knowledge is restricted or not.

The construction of UAP is an optimization process. In the process, a perturbation is added to the training set, and then the perturbation is optimized to make the model output the wrong results as much as possible. In other words, the construction process of UAP is a process of generating a perturbation that can destroy the data distribution of the original training set as much as possible. In the attack process, because the testing samples usually have the same or similar distribution as the training set, after added with UAP, it may deviate from the original distribution, which may make the model output the wrong result.

The process of constructing UAP for the DeepReceiver is shown in Algorithm~\ref{algorithm3}. Specifically, the adversary constructs a new training set $\mathcal {S}_\mathrm{UAP}$ from the original training set $\mathcal {S}$. UAP is added to the training set $\mathcal {S}_\mathrm{UAP}$ to determine whether the BER of the DeepReceiver reaches the preset value. If the preset value is not reached, the UAP is optimized so that the optimized UAP can attack the samples in the training set that were not successfully attacked. In addition, considering the physical feasibility of UAP transmission, the power and PAPR restrictions are considered in the process of constructing UAP. 

\begin{algorithm}[htbp]
\caption{UAP attack method}
\begin{algorithmic}[1]
\renewcommand{\algorithmicrequire}{\textbf{Input:}}
\REQUIRE Training set $\mathcal {S}_\mathrm{UAP}$, DeepReceiver $\mathcal F$, and desired BER $\lambda$.

\renewcommand{\algorithmicrequire}{\textbf{Restrictions:}}
\REQUIRE Power restrictions $\varepsilon$, PAPR restriction $\beta$.

\renewcommand{\algorithmicensure}{\textbf{Output:}}
\ENSURE Universal adversarial perturbation $\delta _{\mathrm{UAP}} ( n )$.

\STATE Initialize $\delta _{\mathrm{UAP}} ( n ) = 0$.

\WHILE{$\operatorname { BER } \left( \mathcal {S}_\mathrm{UAP}+ \sqrt { \varepsilon } \cdot \delta _{\mathrm{UAP}} ( n ) \right) < \lambda$}
    \FOR{each sample $ s _ { i } ( n ) \in \mathcal S$}
        \STATE $result_ { adv }  = \mathcal F \left( s _ { i } ( n ) + \sqrt { \varepsilon } \cdot \delta _{\mathrm{UAP}} ( n ) \right)$.
        \STATE $result_ { s }  = \mathcal F \left( s _ { i } ( n ) \right)$.
        \IF{$result_{adv} == result_{s}$}
            % \STATE Update $\Delta \delta(n)$ by Eq.~(\ref{formal 5}).
            \STATE Input $\delta _{\mathrm{UAP}} ( n )+\Delta \delta(n)$ and update $\Delta \delta(n)$ by Eq.~(\ref{formal 5}).
            
            \STATE Update the adversarial perturbation:\\
            $\delta _{\mathrm{UAP}} ( n ) = \delta _{\mathrm{UAP}} ( n ) + \Delta \delta(n)$
            % \STATE Update $\Delta \delta(n)$:\\
            % \delta ( n )=\arg \max _ { \delta  ( n ) } \| \mathcal{F} ( r ( n ) + \delta (n)) - \mathcal{F} ( r ( n ) ) \| _ { 1 }, \\ 
            % \text { s.t. } \quad \operatorname {Power} ( \delta ( n ) ) \le \varepsilon, \\ 
            % \quad \quad \quad \operatorname {PAPR} ( \delta ( n ) ) \le \beta ,
            
            % \STATE $ \Delta \delta \leftarrow r,$ \\
            %  $\text { s.t. } \mathcal F \left( s _ { i } ( n ) + \delta _ { U A P } ( n ) + r \right) \neq \mathcal F \left( s _ { i } ( n ) \right), $ \\
            %  $\operatorname{PAPR} \left( \delta _ { U A P } ( n ) + r \right) \le \beta, $\\
            %  $\operatorname{Power} \left( \delta _ { U A P } ( n ) + r \right) \le \varepsilon$,
            
            % \STATE Update the adversarial perturbation:\\
            %       $\delta _ { U A P } ( n ) = \delta _ { U A P } ( n ) + \Delta \delta(n) $.
            \STATE Restrict the PAPR of the adversarial perturbation by Eq.~(\ref{e5mifgsm}).
            \STATE Normalize the power of the adversarial perturbation by Eq.~(\ref{e4fgsm}).
        \ENDIF
    \ENDFOR
\ENDWHILE

\RETURN $\delta _{\mathrm{UAP}} ( n )$
\end{algorithmic}
\label{algorithm3}
\end{algorithm}

\section{Experimental Results\label{experimental}}
% Aiming at the problem that the robustness of the information recovery model based on DNN has not been verified, we carry out experiments on the robustness of the state-of-the-art information recovery model called DeepReceiver through adversarial attack. In this section, we show the attack effects of different attack methods on the DeepReceiver in different scenarios to verify it's robustness. In addition, considering the feasibility of the physical transmission of the transmitter, we explored the restrictions of different power and PAPR on the attack effect. Furthermore, we also consider the influence of some hyper-parameters such as the number of iterations and the number of training sets on the attack effect.

\subsection{Experimental Setup}
\subsubsection{Data Generation and Experimental Platform}
We generate training set and testing set according to the data generation process of DeepReceiver~\cite{zheng2020deepreceiver}. Both the training set and the testing set include the information bit streams and the corresponding IQ signals. The information bit stream is randomly generated and the number of bits in the information bit stream is $M=32$. The bit stream will be channel-encoded into 56 bits using Hamming (7,4) coding scheme, modulated using BPSK, and then shaped by raised cosine filtering with roll-off factor 0.5 to become the transmitted signal. The sampling rate of the received signal is 8 times the symbol rate. Therefore, the data length of the IQ signal is 448, which is used as the input of the DeepReceiver model. In the generated training set, the Eb/N0 of the IQ signal ranges from 0 dB to 8 dB with an interval of 1 dB. The number of data samples for each Eb/N0 is 400,000, so the total number of samples in the training set is 3,600,000. In the testing set, Eb/N0 ranges from 0 dB to 8 dB with an interval of 0.5 dB, and the number of samples per Eb/N0 is 200,000.

The experiments were conducted on a server equipped with Intel XEON 6240 2.6GHz X 18C (CPU), Tesla V100 32GiB (GPU), 16GiB DDR4-RECC 2666MHz (Memory), Ubuntu 16.04 (OS), Python 3.6, Keras-2.2.4, Tensorflow-1.10.0. SGD is used to train the DeepReceiver and the surrogate model. The batch size is 256, the number of epochs is 8 and the initial learning rate is 0.001. After every 2 epochs, the learning rate is reduced to 1/10 of the previous learning rate.
% \begin{table}[!htbp]
% \small
% \centering
% \caption{Statistical table of model training parameters}
% \begin{tabular}{m{0pt} m{2.5cm}<{\centering} m{1.8cm}<{\centering} m{2.5cm}<{\centering}}
%     \hline \cline{1-4} 
%     \rule{0pt}{15pt}    & ~                     & DeepReceiver  & Surrogate model \\ \hline 
%     \rule{0pt}{12pt}    & Batch-size            & 256           & 256              \\ \hline 
%     \rule{0pt}{12pt}    & Initial learning rate & 0.001         & 0.001            \\ \hline 
%     \rule{0pt}{12pt}    & Epoch                 & 8             & 8                \\ \hline 
%     \rule{0pt}{12pt}    & Optimizer             & SGD           & SGD              \\ \hline \cline{1-4}
% \end{tabular}
% \label{tab2}
% \end{table}

% \begin{table}[htbp]
% \small
% \centering
% \caption{Statistical table of model training parameters}
% \begin{tabular}{ccc}\hline
% \diagbox {Parameters}{Model} & DeepReceiver & Equivalent model\\[6pt] \hline
% Batch-size   & 256  & 256   \\[3pt]
% \cline{1-3}
% Initial learning rate  & 0.001  & 0.001   \\[3pt]
% \cline{1-3}
% Epoch        & 8   & 8   \\[3pt]
% \cline{1-3}
% Optimizer    & SGD   & SGD   \\[3pt]
% \cline{1-3}
% \end{tabular}
% \label{tab2}
% \end{table}

\subsubsection{Performance Metrics}
We use BER to measure the attack effect. The higher the BER, the better the attack effect. The BER is defined as
\begin{equation}
\operatorname {BER} = \frac { \sum _ { i = 1 } ^ { N_{adv} } m _ { i } } { N_{adv} \cdot M },
\end{equation}

\noindent where $m_{i}$ represents the number of error bits in the recovered stream corresponding to the $i$-th input adversarial signal, and $N_{adv}$ represents the number of adversarial signals. Furthermore, we use the relative power to measure the power of the generated adversarial perturbations and define the adversarial to signal power ratio (PSR) as
\begin{equation}
\operatorname {PSR} = \frac { \sum _ { n = 1 } ^ { N } |\delta  ( n )| ^ { 2 }/ N } { \sum _ { n = 1 } ^ { N } |s_r  ( n )|^ { 2 } / N },
\end{equation}

\noindent where $s_r(n)$ represents the benign signal component in the received signal $r(n)$.

\subsection{Attack Effects in Different Scenarios}
In this section, we explore the effect of different attack methods on the DeepReceiver in different scenarios, without considering the impact of other factors such as power and PAPR. Therefore, in the experiments in this section, the values of PSR and PAPR are fixed, $-5$ dB and 2 dB, respectively. It should be noted that the PAPR of the adversarial perturbations generated by FGSM is 0 dB, which does not to be restricted. The number of iterations of MI-FGSM and PGD is set to 3.
%For the impact of different powers, PAPR and different iterations of PGD and MI-FGSM attack methods on the attack effect will be discussed in Sections~\ref{POWER} and Sections~\ref{Iteration}.

\subsubsection {TestKnown and ModelKnown}
\begin{figure}[t]
    \centering
    \includegraphics[width=0.4\textwidth]{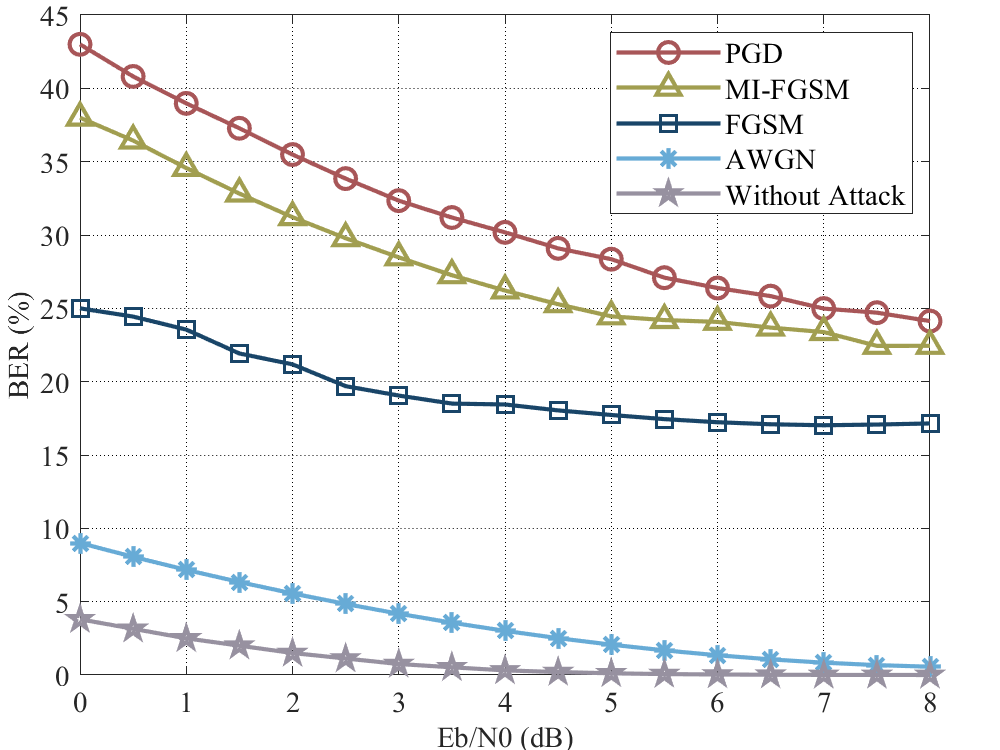}
    \caption{Effects of different attack methods on DeepReceiver.}
    \label{figure3}
\end{figure}

In the scenario of TestKnown and ModelKnown, the attack effects of the adversarial perturbations generated by the three attack methods are shown in Fig.~\ref{figure3}. For comparison, the attack effect of additive white Gaussian noise (AWGN) with the same power on the performance of DeepReceiver is also provided. It can be found that under the same power, FGSM, MI-FGSM and PGD can all achieve attacks on the DeepReceiver and the attack effects are far better than AWGN-based attack. Further observation shows that the attack effect of FGSM attack method is worse than MI-FGSM and PGD, and the attack performance of PGD is the best. However, even the FGSM attack method can increase the BER of DeepReceiver to at least 20$\%$, which shows that DeepReceiver is vulnerable to adversarial attacks.

\subsubsection {TestKnown and ModelUnknown}
\begin{figure}[t]
    \centering
    \includegraphics[width=0.4\textwidth]{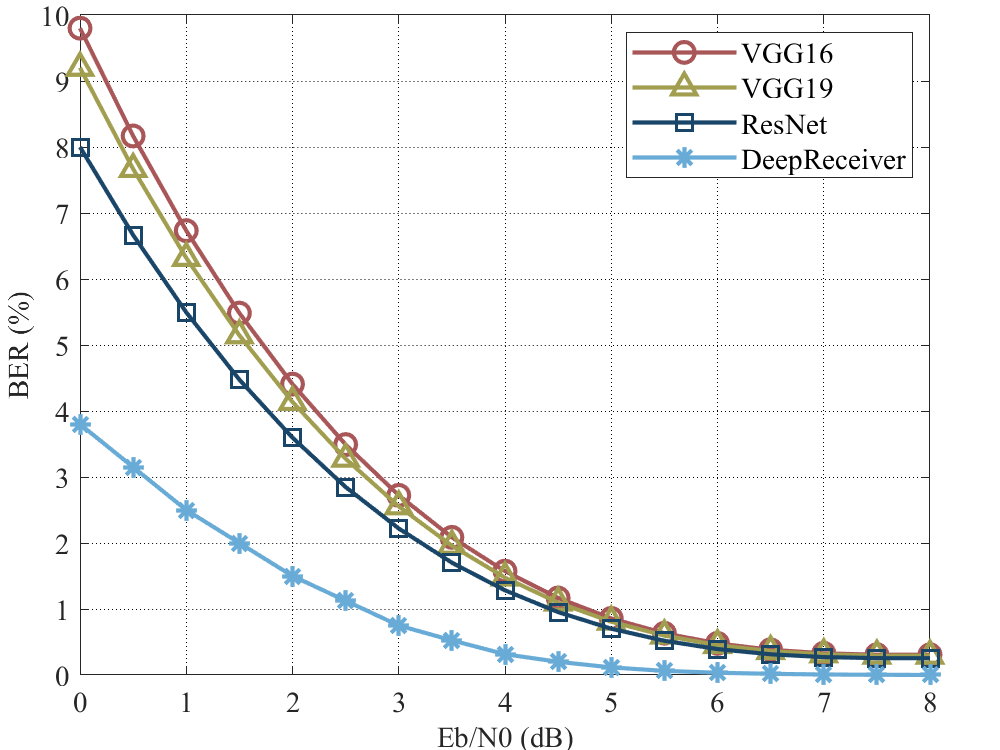}
    \caption{The performance of the constructed surrogate models.}
    \label{figure4}
\end{figure}

\begin{figure*}[htbp]
  \centering
  \subfigure[]{
  \includegraphics[width=0.31\textwidth]{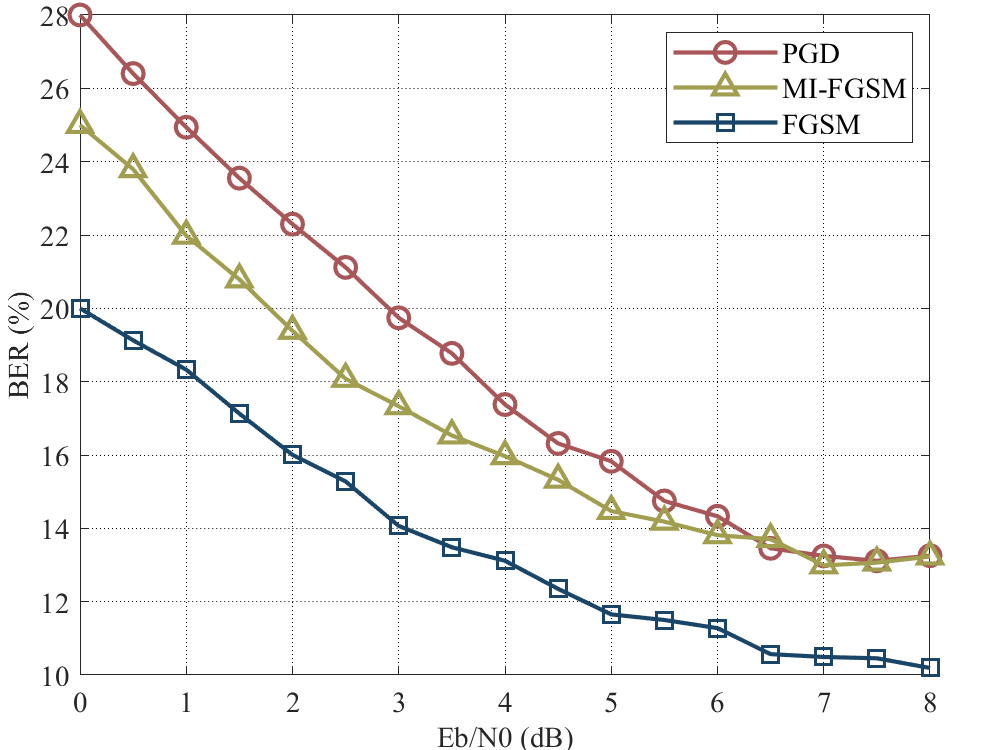}
    \label{fig5a}
    }
  \subfigure[]
  {
  \includegraphics[width=0.31\textwidth]{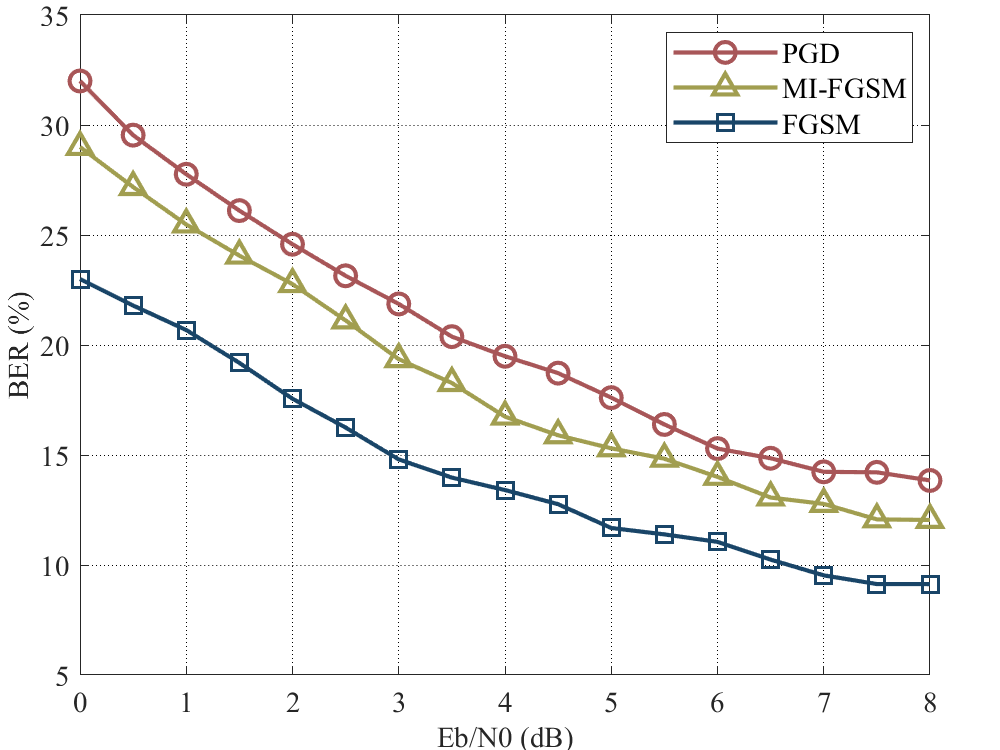}
    \label{fig5b}
    }
  \subfigure[]
  {
  \includegraphics[width=0.31\textwidth]{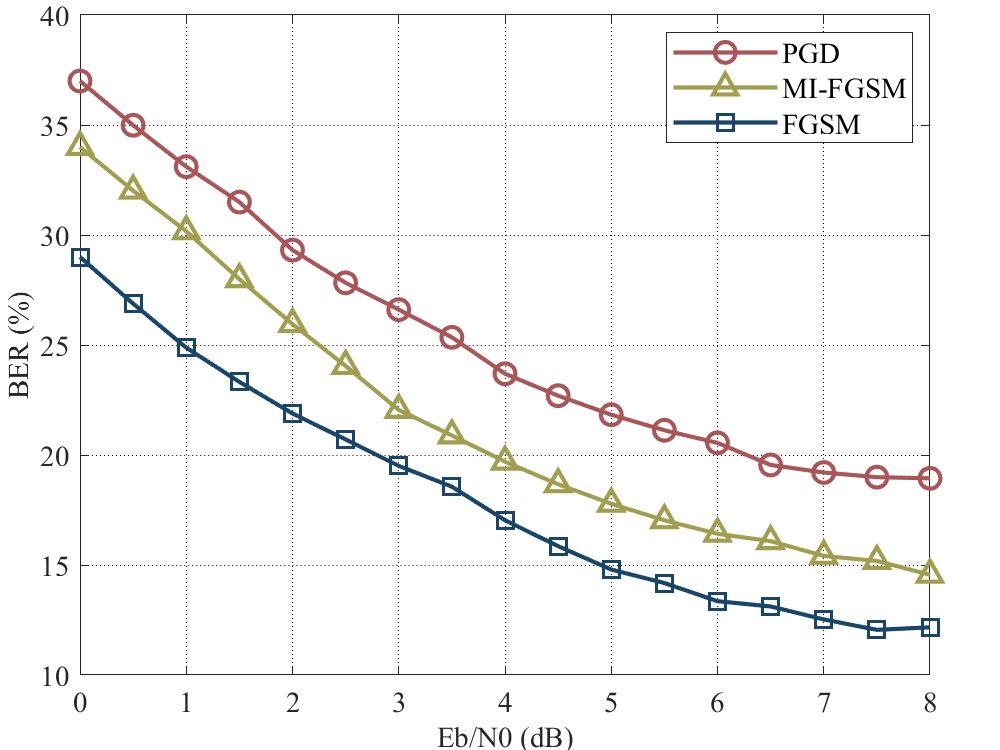}
    \label{fig5c}
    }
  \caption{DeepReceiver attack effects of adversarial perturbations generated on surrogate models. (a) VGG16-based surrogate model, (b) VGG19-based surrogate model, and (c) ResNet-based surrogate model.}
  \label{figure5}
\end{figure*}

In TestKnown and ModelUnknown scenario, the adversary attacks the constructed surrogate model to generate adversarial perturbations which are then used to attack the DeepReceiver. In order to explore the attack effect of the adversarial perturbations generated on the surrogate models of different structures, besides the surrogate model based on Resnet, the surrogate models based on VGG16~\cite{simonyan2014very} and VGG19~\cite{simonyan2014very} are also constructed, whose main structures are different from DeepReceiver. The performance of the constructed surrogate models is show in Fig.~\ref{figure4}. It can be seen that these surrogate models perform worse than DeepReceiver. The attack effects of adversarial perturbations generated on these surrogate models are show in Fig.~\ref{figure5}. It can be found that although the performance of the surrogate models we constructed is not as good as that of the DeepReceiver, the adversarial perturbations generated on these surrogate models can attack DeepReceiver successfully. Furthermore, we can observe that the better the performance of the surrogate model, the better attack effect of adversarial perturbations.

\subsubsection{TestUnknown}
\begin{figure}[htbp]
    \centering
    \includegraphics[width=0.36\textwidth]{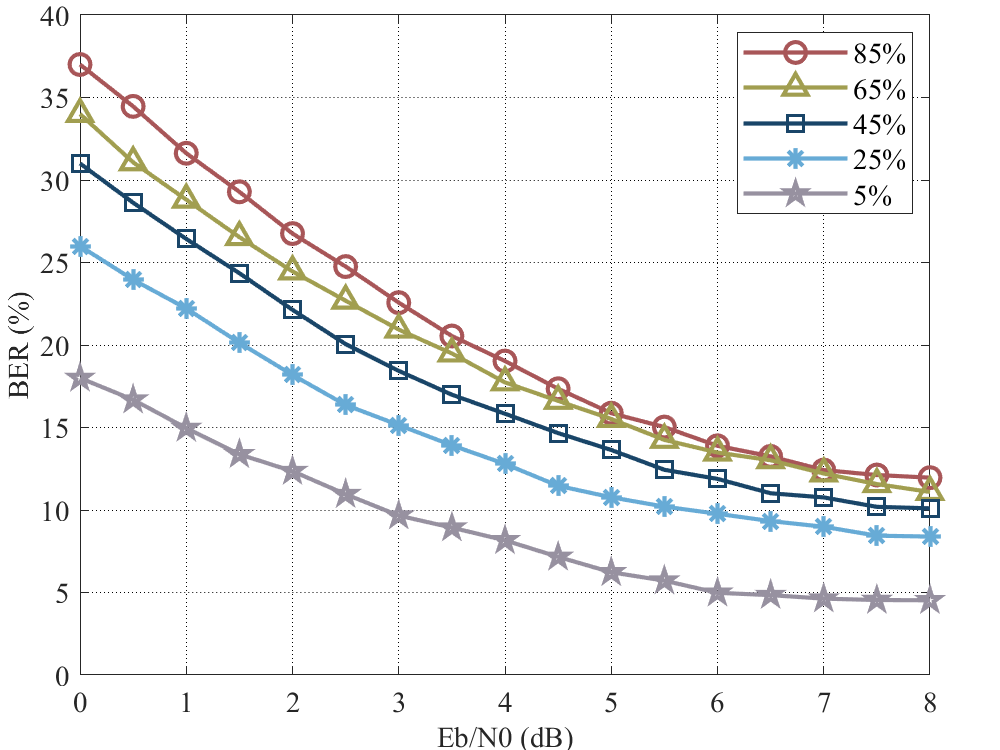}
    \caption{DeepReceiver attack effect of UAP constructed on training sets with different number of samples.}
    \label{figure6}
\end{figure}

In TestUnknown scenario, the adversary can construct a new training set $\mathcal {S}_\mathrm{UAP}$ from the original training set $\mathcal {S}$ and use the constructed training set $\mathcal {S}_\mathrm{UAP}$ to generate UAP to attack DeepReceiver. The number of samples in $\mathcal {S}_\mathrm{UAP}$ will affect attack effect of the UAP. Therefore, we explore the relationship between the number of samples in $\mathcal {S}_\mathrm{UAP}$ and the attack effect of UAP. We assume that the number of samples in $\mathcal {S}_\mathrm{UAP}$ constructed by the adversary accounts for 5$\%$, 25$\%$, 45$\%$, 65$\%$ and 85$\%$ of the original training set $\mathcal {S}$. The attack effect of UAP generated under different number of samples is shown in Fig.~\ref{figure6}. It can be observed that the larger the number of samples in $\mathcal {S}_\mathrm{UAP}$, the better the attack effect. At high Eb/N0, as the number of samples in $\mathcal {S}_\mathrm{UAP}$ increases, the attack performance of UAP will not always be significantly improved. This is because when the number of samples in $\mathcal {S}_\mathrm{UAP}$ reaches a certain value, the gap of data distribution between the constructed training set $\mathcal {S}_\mathrm{UAP}$ and the original training set $\mathcal {S}$ is small, and the increase of number of samples can no longer significantly improve the UAP attack effect. It can also be found that even if the number of samples in the training set constructed by the adversary is about half ($45\%$) of the original training set, the generated UAP can make the BER of DeepReceiver close to 10 $\%$ at high Eb/N0, which shows that DeepReceiver is vulnerable to UAP attacks in TestUnknown scenario.

\subsection{Influence of PSR and PAPR on Attack Effect\label{POWER}}
\begin{figure*}[htbp]
  \centering
  \subfigure[]{
  \includegraphics[width=0.9\textwidth]{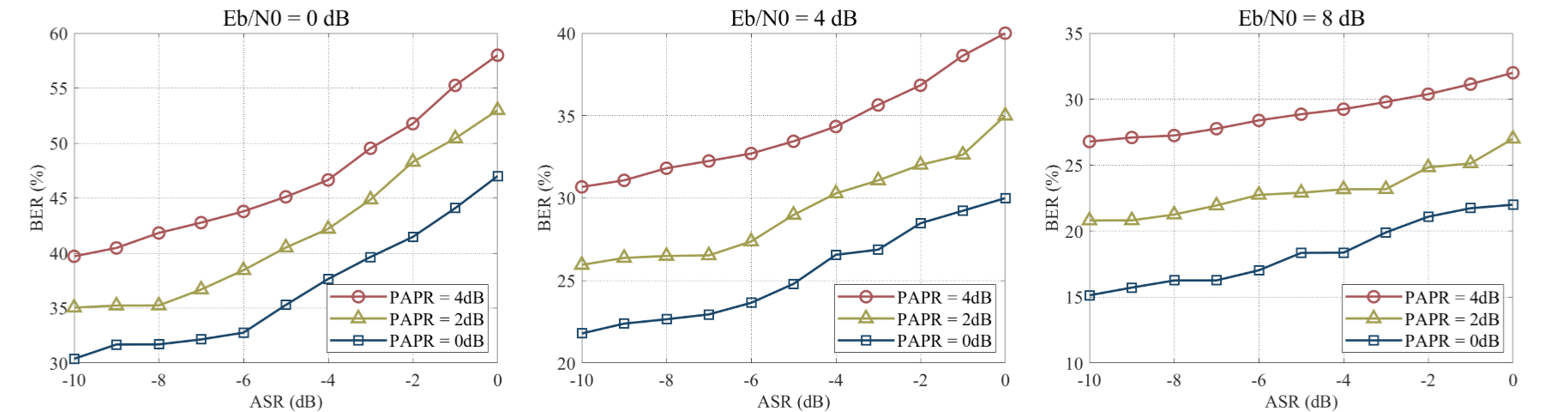}
    }
  \subfigure[]
  {
  \includegraphics[width=0.9\textwidth]{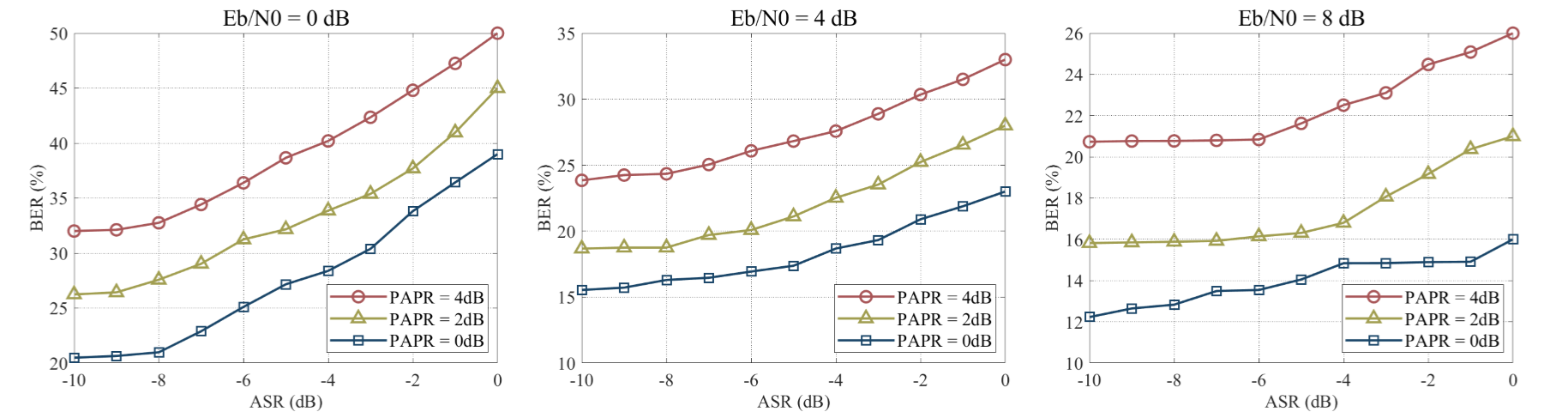}
    }
  \subfigure[]
  {
  \includegraphics[width=0.9\textwidth]{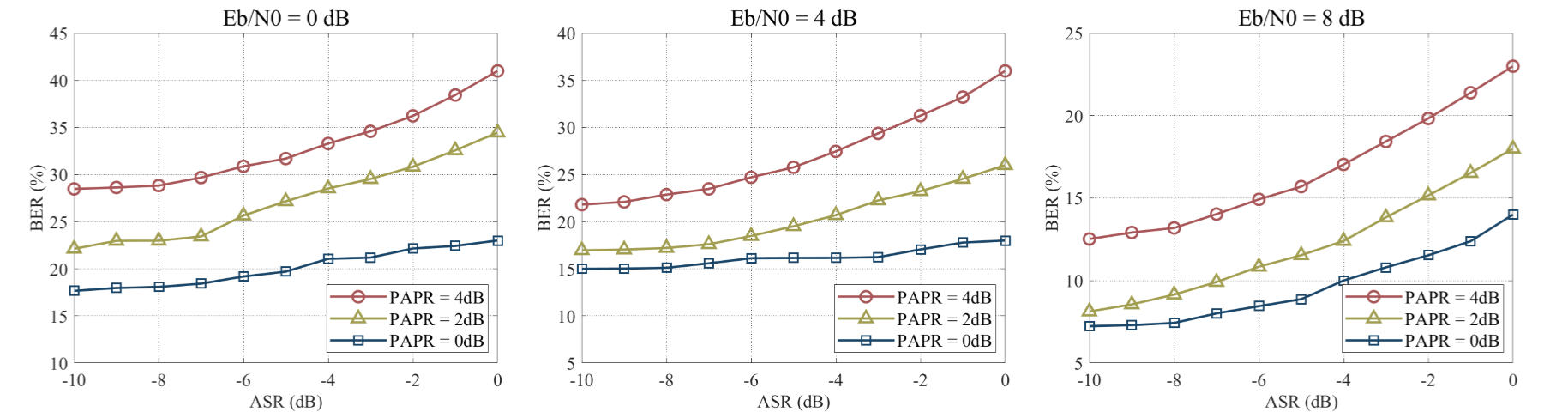}
    }
  \caption{The influence of PSR and PAPR on attack effect. (a) PGD attack method in TestKnown and ModelKnown scenario, (b) Surrogate+PGD attack method in TestKnown and ModelUnknown scenario, and (c) UAP attack method in TestUnknown scenario.}
  \label{figure7}
\end{figure*}

In this section, we explore the attack effect of adversarial perturbations generated under different PAPR and PSR due to power restrictions. Experiments of all attack scenarios in Table~\ref{table1} are conducted. For simplicity, we only show here the results of PGD, surrogate+PGD, and UAP and leave the results of other attack methods in Section \ref{Appendix} Appendix. The number of iterations of the PGD attack method is set to 3. The structure of the surrogate model is ResNet. For the UAP-based attack method, it is assumed that the number of samples in the training set $\mathcal {S}_\mathrm{UAP}$ constructed by the adversary accounts for $25\%$ of the original training set $\mathcal {S}$. The performance of the three attack methods under different PSR and PAPR are shown in Fig.~\ref{figure7}. It can be seen from Fig.~\ref{figure7} that no matter which attack method is, the greater the PSR, the better the attack effect. This is because large power of adversarial perturbations is beneficial to distort the testing samples. However, the purpose of adversarial attack is to generate adversarial perturbations as subtle as possible to disturb normal operation of DeepReceiver. This has been achieved as can be observed from the results. For instance, when $\mathrm {PSR}=-10$ dB which means the power of the adversarial perturbation is only one tenth of that of the signal, the three adversarial attack methods can make the BER of the DeepReceiver more than $10\%$ when Eb/N0 $\le 4$ dB. When Eb/N0 $=8$ dB, even in the most challenging scenario of TestUnknown, when $\mathrm{PSR}=-10$ dB, the BER of DeepReceiver after UAP attack is higher than $5\%$, far higher than the BER of DeepReceiver without attack ($\operatorname{BER} \approx 10^{-5}$ in this case). 

As for PAPR, it can be observed that the greater the PAPR of the adversarial perturbations, the better the attack effect. This is because large PAPR means the maximum amplitude of the adversarial perturbation is large, and in this case the continuity of the IQ signal in the time dimension is easier to be destroyed with this perturbation. Therefore, it's difficult for DeepReceiver to recover the original information bit stream from this corrupted adversarial signal. However, as discussed earlier, the adversary prefers low PAPR in generating adversarial perturbations. In the extreme case of $\operatorname{PAPR}=0$ dB, as can be observed from the results, the three attack methods can disturb the operation of DeepReceiver and the resulting BER is relatively high. These results further show that DeepReceiver is vulnerable to
adversarial attacks.

 %Since there are a total of 17 Eb/N0 in the test set, it would be a huge task to discuss the impact of power and PAPR on the attack effect in each Eb/N0. Therefore, we select the corresponding Eb/N0 for the best, worst and medium performance of DeepReceiver, and the corresponding Eb/N0 are 8dB, 0dB and 4dB respectively.

\subsection{Influence of Iteration Number on Attack Effect\label{Iteration}}
\begin{figure}[htbp]
    \centering
    \includegraphics[width=0.36\textwidth]{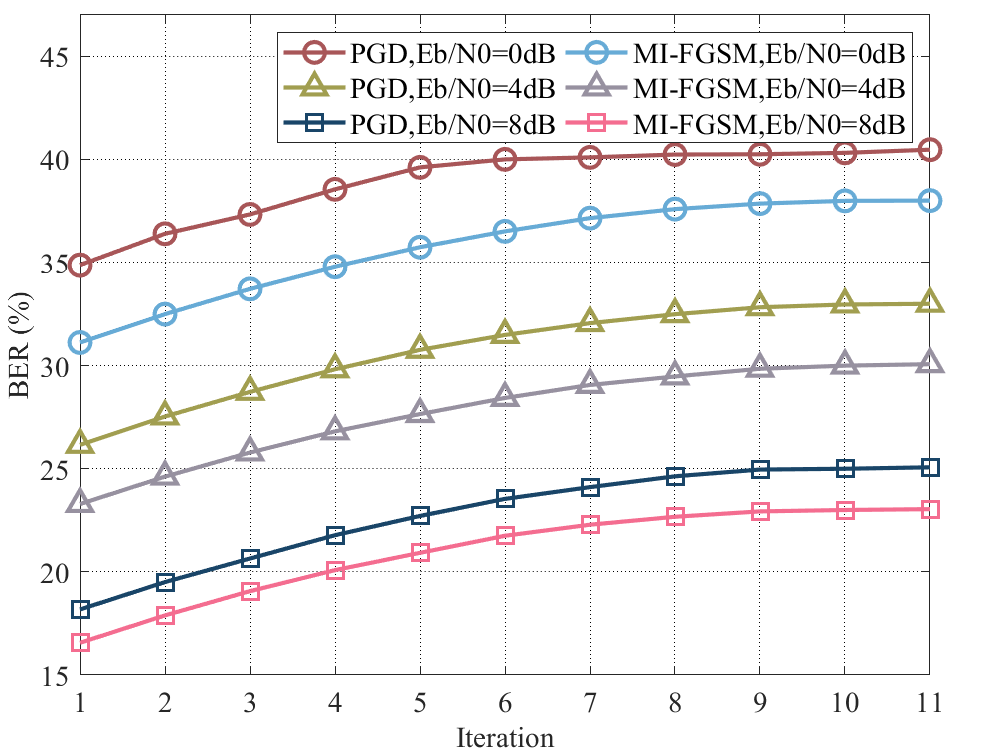}
    \caption{The influence of number of iteration on attack effect.}
    \label{figure8}
\end{figure}

MI-FGSM and PGD are two gradient-based iterative attack methods, and the number of iterations will affect the attack effect. In this section, we explore the impact of the number of iterations in the MI-FGSM and PGD attack methods on the attack effect. In order to only explore the impact of number of iterations on the attack effect, the PSR is fixed to $-5$ dB and the PAPR is fixed to 2 dB. As for Eb/N0, we select the corresponding Eb/N0 for the best, worst and medium performance of DeepReceiver, and the corresponding Eb/N0 are 8 dB, 0 dB and 4 dB respectively. The effect of the number of iterations on the attack effect is shown in Fig.~\ref{figure8}. It can be seen that for the two methods, the number of iteration is higher, the attack effect of adversarial perturbation is better. But when the number of iterations reaches a certain value, the attack effect will no longer improve. This is because when the number of iterations reaches a certain value, the gradient of the loss function is close to 0, and the constructed adversarial perturbation is nearly the optimal solution in the search space.

\begin{figure*}[t]
  \centering
  \subfigure[]
  {
  \includegraphics[width=0.9\textwidth]{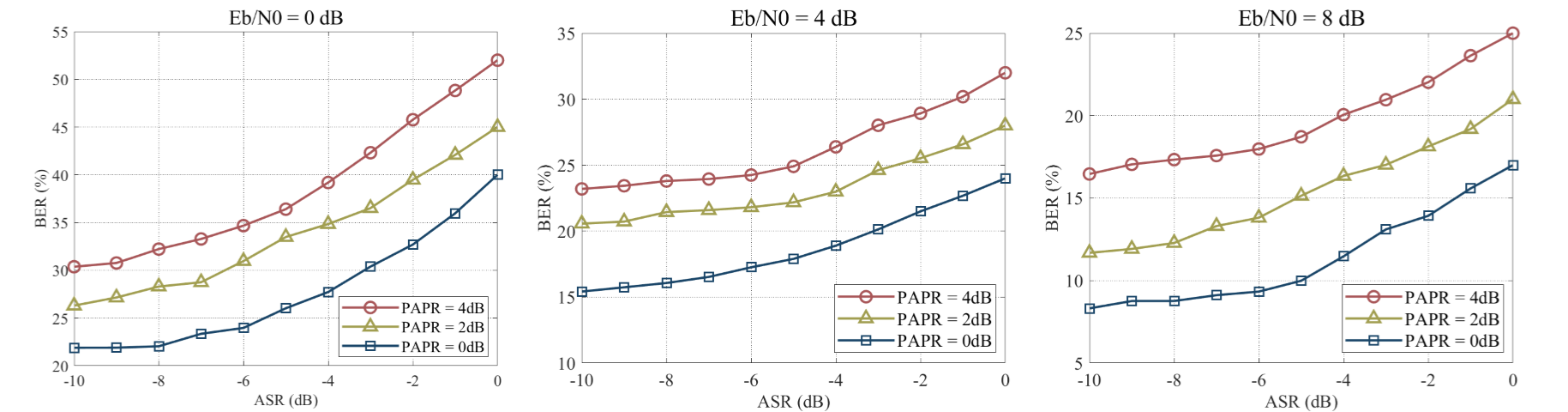}
    }

   \subfigure[]
  {
  \includegraphics[width=0.9\textwidth]{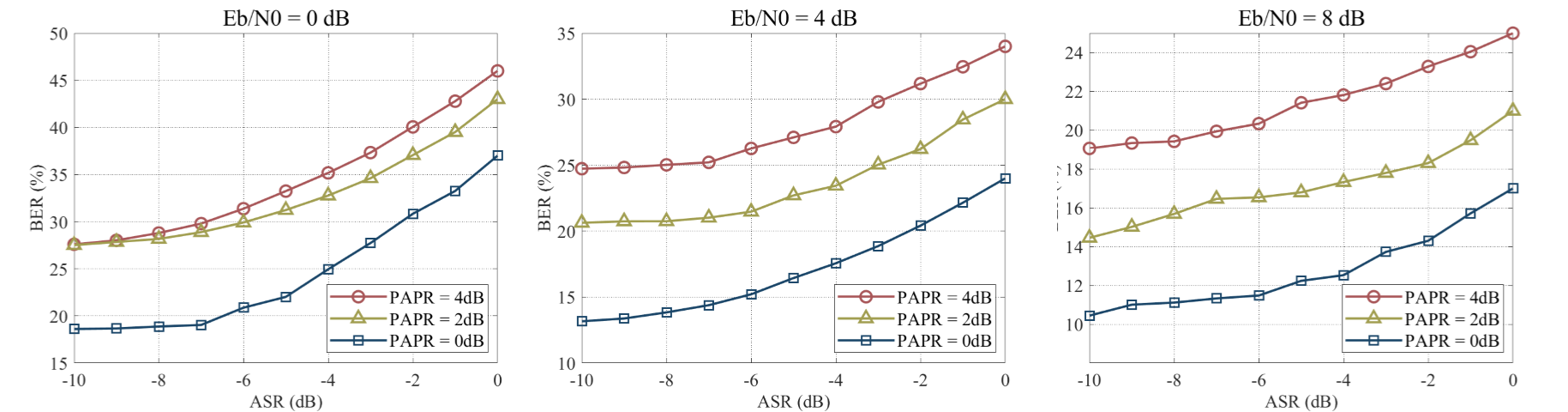}
    }
  \caption{Influence of PSR and PAPR on attack effect. (a) MI-FGSM attack method in TestKnown and ModelKnown scenario and (b) surrogate+MI-FGSM attack method in TestKnown and ModelUnknown scenario.}
  \label{figure9}
\end{figure*}

\section{Conclusion\label{conclusion}}
Aiming at the problem that the robustness of the DL-based information recovery model has not been investigated, we have tested the SOTA information recovery model named DeepReceiver to study the robustness of the information recovery model. We have systematically explored the possible security vulnerabilities of the DeepReceiver in multiple scenarios. According to our experimental results, it can be found that the DeepReceiver model has security vulnerabilities in these scenarios. Even in the scenario of TestUnknown (and ModelUnknown), the adversary can attack the DeepReceiver and increase its BER above 10$\%$ when adequate training samples are available and PSR and PAPR are fixed to $-5$ dB and 2 dB. We have further explored the influence of ISR, PAPR and the hyperparameter in the attack method on the attack effect. From these experiments, it can be found that the DeepReceiver is vulnerable to adversarial attacks even with low ISR (power) and limited PAPR. After our verification, it can be found that DeepReceiver has security vulnerabilities in the face of the proposed AIR. Defense measures should be taken to enhance the security of DeepReceiver.

%\appendices
\section*{Appendix\label{Appendix}}
The attack effects of the other four attack methods in Table~\ref{table1} under different PSR and PAPR are shown in Fig.~\ref{figure9}. The number of iterations of the MI-FGSM attack method is set to 3. The structure of the surrogate model is ResNet. It should be noted that the PAPR of adversarial perturbations generated by the FGSM attack method is 0 dB, which is the minimum value, so the impact of the PAPR restriction on the attack effect of the FGSM attack method is not considered. It can be found that the influence of PSR and PAPR on the other four attack methods is similar to the experimental results in Section~\ref{POWER}.

% you can choose not to have a title for an appendix
% if you want by leaving the argument blank
%\section{a}
%Appendix two text goes here.

% use section* for acknowledgment
%\section*{Acknowledgment}
%The authors would like to thank...

% Can use something like this to put references on a page
% by themselves when using endfloat and the captionsoff option.
\ifCLASSOPTIONcaptionsoff
  \newpage
\fi

% trigger a \newpage just before the given reference
% number - used to balance the columns on the last page
% adjust value as needed - may need to be readjusted if
% the document is modified later
%\IEEEtriggeratref{8}
% The "triggered" command can be changed if desired:
%\IEEEtriggercmd{\enlargethispage{-5in}}

% references section

% can use a bibliography generated by BibTeX as a .bbl file
% BibTeX documentation can be easily obtained at:
% http://mirror.ctan.org/biblio/bibtex/contrib/doc/
% The IEEEtran BibTeX style support page is at:
% http://www.michaelshell.org/tex/ieeetran/bibtex/
%\bibliographystyle{IEEEtran}
% argument is your BibTeX string definitions and bibliography database(s)
%\bibliography{IEEEabrv,../bib/paper}
%
% <OR> manually copy in the resultant .bbl file
% set second argument of \begin to the number of references
% (used to reserve space for the reference number labels box)

\bibliography{bare_jrnl}

% Generated by IEEEtran.bst, version: 1.14 (2015/08/26)
\begin{thebibliography}{10}
\providecommand{\url}[1]{#1}
\csname url@samestyle\endcsname
\providecommand{\newblock}{\relax}
\providecommand{\bibinfo}[2]{#2}
\providecommand{\BIBentrySTDinterwordspacing}{\spaceskip=0pt\relax}
\providecommand{\BIBentryALTinterwordstretchfactor}{4}
\providecommand{\BIBentryALTinterwordspacing}{\spaceskip=\fontdimen2\font plus
\BIBentryALTinterwordstretchfactor\fontdimen3\font minus
  \fontdimen4\font\relax}
\providecommand{\BIBforeignlanguage}[2]{{%
\expandafter\ifx\csname l@#1\endcsname\relax
\typeout{** WARNING: IEEEtran.bst: No hyphenation pattern has been}%
\typeout{** loaded for the language `#1'. Using the pattern for}%
\typeout{** the default language instead.}%
\else
\language=\csname l@#1\endcsname
\fi
#2}}
\providecommand{\BIBdecl}{\relax}
\BIBdecl

\bibitem{khan2021role}
S.~K. Khan, U.~Naseem, H.~Siraj, M.~I. Razzak, and M.~Imran, ``The role of
  unmanned aerial vehicles and mmwave in 5g: recent advances and challenges,''
  \emph{Transactions on emerging telecommunications technologies}, pp. 1--18,
  2021.

\bibitem{duan2020emerging}
W.~Duan, J.~Gu, M.~Wen, G.~Zhang, Y.~Ji, and S.~Mumtaz, ``Emerging technologies
  for 5g-iov networks: Applications, trends and opportunities,'' \emph{IEEE
  Network}, vol.~34, no.~5, pp. 283--289, 2020.

\bibitem{gao2019mobile}
Y.~Gao, Z.~Deng, Y.~Zhang, S.~Sun, and Z.~Li, ``Mobile phone passive
  positioning through the detection of uplink signal,'' in \emph{2019 IEEE
  International Conference on Smart Internet of Things (SmartIoT)}.\hskip 1em
  plus 0.5em minus 0.4em\relax IEEE, 2019, pp. 367--371.

\bibitem{nguyen2022Iot}
D.~C. Nguyen, M.~Ding, P.~N. Pathirana, A.~Seneviratne, J.~Li, D.~Niyato,
  O.~Dobre, and H.~V. Poor, ``\BIBforeignlanguage{English}{6g internet of
  things: A comprehensive survey},'' \emph{\BIBforeignlanguage{English}{IEEE
  internet of things journal}}, vol.~9, no.~1, pp. 359--383, 2022.

\bibitem{ghosh20195G}
A.~Ghosh, A.~Maeder, M.~Baker, and D.~Chandramouli,
  ``\BIBforeignlanguage{English}{5g evolution: A view on 5g cellular technology
  beyond 3gpp release 15},'' \emph{\BIBforeignlanguage{English}{IEEE access}},
  vol.~7, pp. 127\,639--127\,651, 2019.

\bibitem{harris2000structure}
F.~Harris and C.~Dick, ``On structure and implementation of algorithms for
  carrier and symbol synchronization in software defined radios,'' in
  \emph{2000 10th European Signal Processing Conference}.\hskip 1em plus 0.5em
  minus 0.4em\relax IEEE, 2000, pp. 1--4.

\bibitem{van1995channel}
J.-J. Van De~Beek, O.~Edfors, M.~Sandell, S.~K. Wilson, and P.~O. Borjesson,
  ``On channel estimation in ofdm systems,'' in \emph{1995 IEEE 45th Vehicular
  Technology Conference. Countdown to the Wireless Twenty-First Century},
  vol.~2.\hskip 1em plus 0.5em minus 0.4em\relax IEEE, 1995, pp. 815--819.

\bibitem{chen1993clustering}
S.~Chen, B.~Mulgrew, and P.~M. Grant, ``A clustering technique for digital
  communications channel equalization using radial basis function networks,''
  \emph{IEEE Transactions on neural networks}, vol.~4, no.~4, pp. 570--590,
  1993.

\bibitem{hoeher1999turbo}
P.~Hoeher and J.~Lodge, ``Turbo dpsk: iterative differential psk demodulation
  and channel decoding,'' \emph{IEEE Transactions on Communications}, vol.~47,
  no.~6, pp. 837--843, 1999.

\bibitem{bai2019deep}
Q.~Bai, J.~Wang, Y.~Zhang, and J.~Song, ``Deep learning-based channel
  estimation algorithm over time selective fading channels,'' \emph{IEEE
  Transactions on Cognitive Communications and Networking}, vol.~6, no.~1, pp.
  125--134, 2019.

\bibitem{yang2019deep}
Y.~Yang, F.~Gao, X.~Ma, and S.~Zhang, ``Deep learning-based channel estimation
  for doubly selective fading channels,'' \emph{IEEE Access}, vol.~7, pp.
  36\,579--36\,589, 2019.

\bibitem{mao2019roemnet}
H.~Mao, H.~Lu, Y.~Lu, and D.~Zhu, ``Roemnet: Robust meta learning based channel
  estimation in ofdm systems,'' in \emph{ICC 2019-2019 IEEE International
  Conference on Communications (ICC)}.\hskip 1em plus 0.5em minus 0.4em\relax
  IEEE, 2019, pp. 1--6.

\bibitem{soltani2019deep}
M.~Soltani, V.~Pourahmadi, A.~Mirzaei, and H.~Sheikhzadeh, ``Deep
  learning-based channel estimation,'' \emph{IEEE Communications Letters},
  vol.~23, no.~4, pp. 652--655, 2019.

\bibitem{he2018deep}
H.~He, C.-K. Wen, S.~Jin, and G.~Y. Li, ``Deep learning-based channel
  estimation for beamspace mmwave massive mimo systems,'' \emph{IEEE Wireless
  Communications Letters}, vol.~7, no.~5, pp. 852--855, 2018.

\bibitem{mehrabi2019decision}
M.~Mehrabi, M.~Mohammadkarimi, M.~Ardakani, and Y.~Jing, ``Decision directed
  channel estimation based on deep neural network $ k $-step predictor for mimo
  communications in 5g,'' \emph{IEEE Journal on Selected Areas in
  Communications}, vol.~37, no.~11, pp. 2443--2456, 2019.

\bibitem{zamanipour2019survey}
M.~Zamanipour, ``A survey on deep-learning based techniques for modeling and
  estimation of massivemimo channels,'' \emph{arXiv preprint arXiv:1910.03390},
  2019.

\bibitem{zhang2020machine}
L.~Zhang and L.-L. Yang, ``Machine learning for joint channel equalization and
  signal detection,'' \emph{Machine Learning for Future Wireless
  Communications}, pp. 213--241, 2020.

\bibitem{fang2017deep}
L.~Fang and L.~Wu, ``Deep learning detection method for signal demodulation in
  short range multipath channel,'' in \emph{2017 IEEE 2nd International
  Conference on Opto-Electronic Information Processing (ICOIP)}.\hskip 1em plus
  0.5em minus 0.4em\relax IEEE, 2017, pp. 16--20.

\bibitem{shental2019machine}
O.~Shental and J.~Hoydis, ``Machine llrning: Learning to softly demodulate,''
  in \emph{2019 IEEE Globecom Workshops (GC Wkshps)}.\hskip 1em plus 0.5em
  minus 0.4em\relax IEEE, 2019, pp. 1--7.

\bibitem{nachmani2016learning}
E.~Nachmani, Y.~Be'ery, and D.~Burshtein, ``Learning to decode linear codes
  using deep learning,'' in \emph{2016 54th Annual Allerton Conference on
  Communication, Control, and Computing (Allerton)}.\hskip 1em plus 0.5em minus
  0.4em\relax IEEE, 2016, pp. 341--346.

\bibitem{nachmani2018near}
E.~Nachmani, Y.~Bachar, E.~Marciano, D.~Burshtein, and Y.~Be'ery, ``Near
  maximum likelihood decoding with deep learning,'' \emph{arXiv preprint
  arXiv:1801.02726}, 2018.

\bibitem{nachmani2018deep}
E.~Nachmani, E.~Marciano, L.~Lugosch, W.~J. Gross, D.~Burshtein, and Y.~Be'ery,
  ``Deep learning methods for improved decoding of linear codes,'' \emph{IEEE
  Journal of Selected Topics in Signal Processing}, vol.~12, no.~1, pp.
  119--131, 2018.

\bibitem{xu2018polar}
W.~Xu, X.~You, C.~Zhang, and Y.~Be’ery, ``Polar decoding on sparse graphs
  with deep learning,'' in \emph{2018 52nd Asilomar Conference on Signals,
  Systems, and Computers}.\hskip 1em plus 0.5em minus 0.4em\relax IEEE, 2018,
  pp. 599--603.

\bibitem{doan2019neural}
N.~Doan, S.~A. Hashemi, E.~N. Mambou, T.~Tonnellier, and W.~J. Gross, ``Neural
  belief propagation decoding of crc-polar concatenated codes,'' in \emph{ICC
  2019-2019 IEEE International Conference on Communications (ICC)}.\hskip 1em
  plus 0.5em minus 0.4em\relax IEEE, 2019, pp. 1--6.

\bibitem{zheng2020deepreceiver}
S.~Zheng, S.~Chen, and X.~Yang, ``Deepreceiver: a deep learning-based
  intelligent receiver for wireless communications in the physical layer,''
  \emph{IEEE Transactions on Cognitive Communications and Networking}, vol.~7,
  no.~1, pp. 5--20, 2020.

\bibitem{szegedy2013intriguing}
C.~Szegedy, W.~Zaremba, I.~Sutskever, J.~Bruna, D.~Erhan, I.~Goodfellow, and
  R.~Fergus, ``Intriguing properties of neural networks,'' \emph{arXiv preprint
  arXiv:1312.6199}, 2013.

\bibitem{kim2020over}
B.~Kim, Y.~E. Sagduyu, K.~Davaslioglu, T.~Erpek, and S.~Ulukus, ``Over-the-air
  adversarial attacks on deep learning based modulation classifier over
  wireless channels,'' in \emph{2020 54th Annual Conference on Information
  Sciences and Systems (CISS)}.\hskip 1em plus 0.5em minus 0.4em\relax IEEE,
  2020, pp. 1--6.

\bibitem{lin2020threats}
Y.~Lin, H.~Zhao, Y.~Tu, S.~Mao, and Z.~Dou, ``Threats of adversarial attacks in
  dnn-based modulation recognition,'' in \emph{IEEE INFOCOM 2020-IEEE
  Conference on Computer Communications}.\hskip 1em plus 0.5em minus
  0.4em\relax IEEE, 2020, pp. 2469--2478.

\bibitem{hameed2020best}
M.~Z. Hameed, A.~Gy{\"o}rgy, and D.~G{\"u}nd{\"u}z, ``The best defense is a
  good offense: Adversarial attacks to avoid modulation detection,'' \emph{IEEE
  Transactions on Information Forensics and Security}, vol.~16, pp. 1074--1087,
  2020.

\bibitem{flowers2019evaluating}
B.~Flowers, R.~M. Buehrer, and W.~C. Headley, ``Evaluating adversarial evasion
  attacks in the context of wireless communications,'' \emph{IEEE Transactions
  on Information Forensics and Security}, vol.~15, pp. 1102--1113, 2019.

\bibitem{kokalj2019adversarial}
S.~Kokalj-Filipovic and R.~Miller, ``Adversarial examples in rf deep learning:
  detection of the attack and its physical robustness,'' \emph{arXiv preprint
  arXiv:1902.06044}, 2019.

\bibitem{usama2019adversarial}
M.~Usama, M.~Asim, J.~Qadir, A.~Al-Fuqaha, and M.~A. Imran, ``Adversarial
  machine learning attack on modulation classification,'' in \emph{2019
  UK/China Emerging Technologies (UCET)}.\hskip 1em plus 0.5em minus
  0.4em\relax IEEE, 2019, pp. 1--4.

\bibitem{kim2020channel}
B.~Kim, Y.~E. Sagduyu, T.~Erpek, K.~Davaslioglu, and S.~Ulukus, ``Channel
  effects on surrogate models of adversarial attacks against wireless signal
  classifiers,'' \emph{arXiv preprint arXiv:2012.02160}, 2020.

\bibitem{kim2020adversarial}
------, ``Adversarial attacks with multiple antennas against deep
  learning-based modulation classifiers,'' in \emph{2020 IEEE Globecom
  Workshops (GC Wkshps}.\hskip 1em plus 0.5em minus 0.4em\relax IEEE, 2020, pp.
  1--6.

\bibitem{sadeghi2018adversarial}
M.~Sadeghi and E.~G. Larsson, ``Adversarial attacks on deep-learning based
  radio signal classification,'' \emph{IEEE Wireless Communications Letters},
  vol.~8, no.~1, pp. 213--216, 2018.

\bibitem{shi2018spectrum}
Y.~Shi, T.~Erpek, Y.~E. Sagduyu, and J.~H. Li, ``Spectrum data poisoning with
  adversarial deep learning,'' in \emph{MILCOM 2018-2018 IEEE Military
  Communications Conference (MILCOM)}.\hskip 1em plus 0.5em minus 0.4em\relax
  IEEE, 2018, pp. 407--412.

\bibitem{sagduyu2019iot}
Y.~E. Sagduyu, Y.~Shi, and T.~Erpek, ``Iot network security from the
  perspective of adversarial deep learning,'' in \emph{2019 16th Annual IEEE
  International Conference on Sensing, Communication, and Networking
  (SECON)}.\hskip 1em plus 0.5em minus 0.4em\relax IEEE, 2019, pp. 1--9.

\bibitem{sagduyu2019adversarial}
Y.~Sagduyu, Y.~Shi, and T.~Erpek, ``Adversarial deep learning for over-the-air
  spectrum poisoning attacks,'' \emph{IEEE Transactions on Mobile Computing},
  2019.

\bibitem{Haghighat2015}
M.~Haghighat and S.~Sadough, ``\BIBforeignlanguage{English}{Smart primary user
  emulation in cognitive radio networks: defence strategies against radio-aware
  attacks and robust spectrum sensing},''
  \emph{\BIBforeignlanguage{English}{Transactions on emerging
  telecommunications technologies}}, vol.~26, no.~9, pp. 1154--1164, 2015.

\bibitem{zheng2021primary}
S.~Zheng, L.~Ye, X.~Wang, J.~Chen, H.~Zhou, C.~Lou, Z.~Zhao, and X.~Yang,
  ``Primary user adversarial attacks on deep learning-based spectrum sensing
  and the defense method,'' \emph{China Communications}, vol.~18, no.~12,
  p.~14, 2021.

\bibitem{goodfellow2014explaining}
I.~J. Goodfellow, J.~Shlens, and C.~Szegedy, ``Explaining and harnessing
  adversarial examples,'' \emph{arXiv preprint arXiv:1412.6572}, 2014.

\bibitem{dong2018boosting}
Y.~Dong, F.~Liao, T.~Pang, H.~Su, J.~Zhu, X.~Hu, and J.~Li, ``Boosting
  adversarial attacks with momentum,'' in \emph{Proceedings of the IEEE
  conference on computer vision and pattern recognition}, 2018, pp. 9185--9193.

\bibitem{madry2017towards}
A.~Madry, A.~Makelov, L.~Schmidt, D.~Tsipras, and A.~Vladu, ``Towards deep
  learning models resistant to adversarial attacks,'' \emph{arXiv preprint
  arXiv:1706.06083}, 2017.

\bibitem{moosavi2017universal}
S.-M. Moosavi-Dezfooli, A.~Fawzi, O.~Fawzi, and P.~Frossard, ``Universal
  adversarial perturbations,'' in \emph{Proceedings of the IEEE conference on
  computer vision and pattern recognition}, 2017, pp. 1765--1773.

\bibitem{he2016deep}
K.~He, X.~Zhang, S.~Ren, and J.~Sun, ``Deep residual learning for image
  recognition,'' in \emph{Proceedings of the IEEE conference on computer vision
  and pattern recognition}, 2016, pp. 770--778.

\bibitem{simonyan2014very}
K.~Simonyan and A.~Zisserman, ``Very deep convolutional networks for
  large-scale image recognition,'' \emph{arXiv preprint arXiv:1409.1556}, 2014.

\end{thebibliography}
\end{document}